\documentclass[aps,pre,reprint,superscriptaddress,showpacs]{revtex4-1}
\pdfoutput=1
\usepackage{graphicx}
\usepackage{color}
\usepackage{amsfonts}
\usepackage{bbm}
\usepackage{amsmath}
\usepackage[letterpaper]{hyperref}
\newcommand{\etal}{{\it{et al.}}}
\begin{document}


\title{A Rich Spectrum of Neural Field Dynamics in the Presence of
\\Short-Term Synaptic Depression}

\author{He Wang}
\author{Kin Lam}
\author{C. C. Alan Fung}
\author{K. Y. Michael Wong}
\affiliation{Department of Physics, Hong Kong 
University of Science and Technology, Hong Kong, China}
\author{Si Wu}
\affiliation{State Key Laboratory of Cognitive Neuroscience and Learning,
IDG/McGovern Institute for Brain Research, 
Beijing Normal University, Beijing 100875, China}

\date{\today}

\begin{abstract}
In continuous attractor neural networks (CANNs), spatially 
continuous information such as orientation, head direction, 
and spatial location is represented by Gaussian-like tuning curves 
that can be displaced continuously in the space of the preferred stimuli 
of the neurons. 
We investigate how short-term synaptic depression (STD)
can reshape the intrinsic dynamics of the CANN model and 
its responses to a single static input.
In particular, CANNs with STD can support various complex firing patterns
and chaotic behaviors.  These chaotic behaviors have the 
potential to encode various stimuli in the neuronal system.
\end{abstract}

\pacs{87.85.dq,
05.45.-a, 
87.19.ll}

\maketitle

\section{\label{sec:intro}introduction}

To understand the brain's operation, it is important to consider the range of 
firing patterns and their conditions of occurrence, which will shed light on 
how information is encoded in the brain \cite{Amit1992}. 
In the processing of continuous information such as object orientation and 
spatial location, firing patterns are found to be localized in the space of 
preferred stimuli of the neurons, normally taking up a Gaussian-like profile 
\cite{Amari1977, Ben-Yishai1995}. Thus an interesting question is 
whether these profiles are stable in time and in space and, 
if not, what other dynamical states will replace them.

Gaussian-like profiles play an important role in both experiments
and theory. The tuning curves, i.e., the functional dependence of 
the neuronal response on the inputs and the preferred stimuli of 
the neurons, are observed as Gaussian-like profiles in various 
animal experiments. For example, head direction cells in anterior 
thalamus and postsubiculum were found in rodents 
\cite{Blair1995, Zhang1996, Taube1998}. Their tuning curves are 
Gaussian-like and centered at their preferred head directions. Place 
cells in hippocampus are another 
example \cite{Okeefe1993, Samsonovich1997}. Place cells' activities 
are observed to depend on the location of the animal within 
the environment. The tuning curve is also a Gaussian-like function. 
Moving direction cells in middle temporal (MT or V5) cortex in 
macaques are found to be selective to object moving directions, 
which also have Gaussian-like tuning curves \cite{Maunsell1983, Treue2000}.

In neural field models processing continuous information, Gaussian-like 
tuning curves are steady states of the network dynamics, and remain 
stable when their positions are displaced in the space of the preferred 
stimuli of the neurons. These neural field models are called continuous 
attractor neural networks (CANNs), since the Gaussian-like tuning 
curves are attractors of the network dynamics. Recent evidence 
supporting the existence of continuous attractors was reported by 
Wimmer \etal, who discovered activity anticorrelations on the opposite 
sides of tuning curves as predicted by the CANN model in the 
prefrontal cortex of monkeys \cite{Pouget1998, Wu2008, Wimmer2014}.

The ability to support these bump attractors is effected by 
couplings between neurons in CANNs. However, in reality, couplings between 
neurons are not quenched. They depend on firing histories of 
presynaptic neurons. Tsodyks \etal\ found that synaptic efficacy 
decreases with firing history \cite{Tsodyks1997, Tsodyks1998}. 
Furthermore, they proposed that this decline in synaptic efficacy is due to 
the slow dynamics of the recovery process of neurotransmitters. The 
recovery of neurotransmitters is of the order of 100 ms. This short-term 
decline in synaptic efficacy is called short-term synaptic depression (STD).

\begin{figure}[ht]
\includegraphics[width=8.6cm]{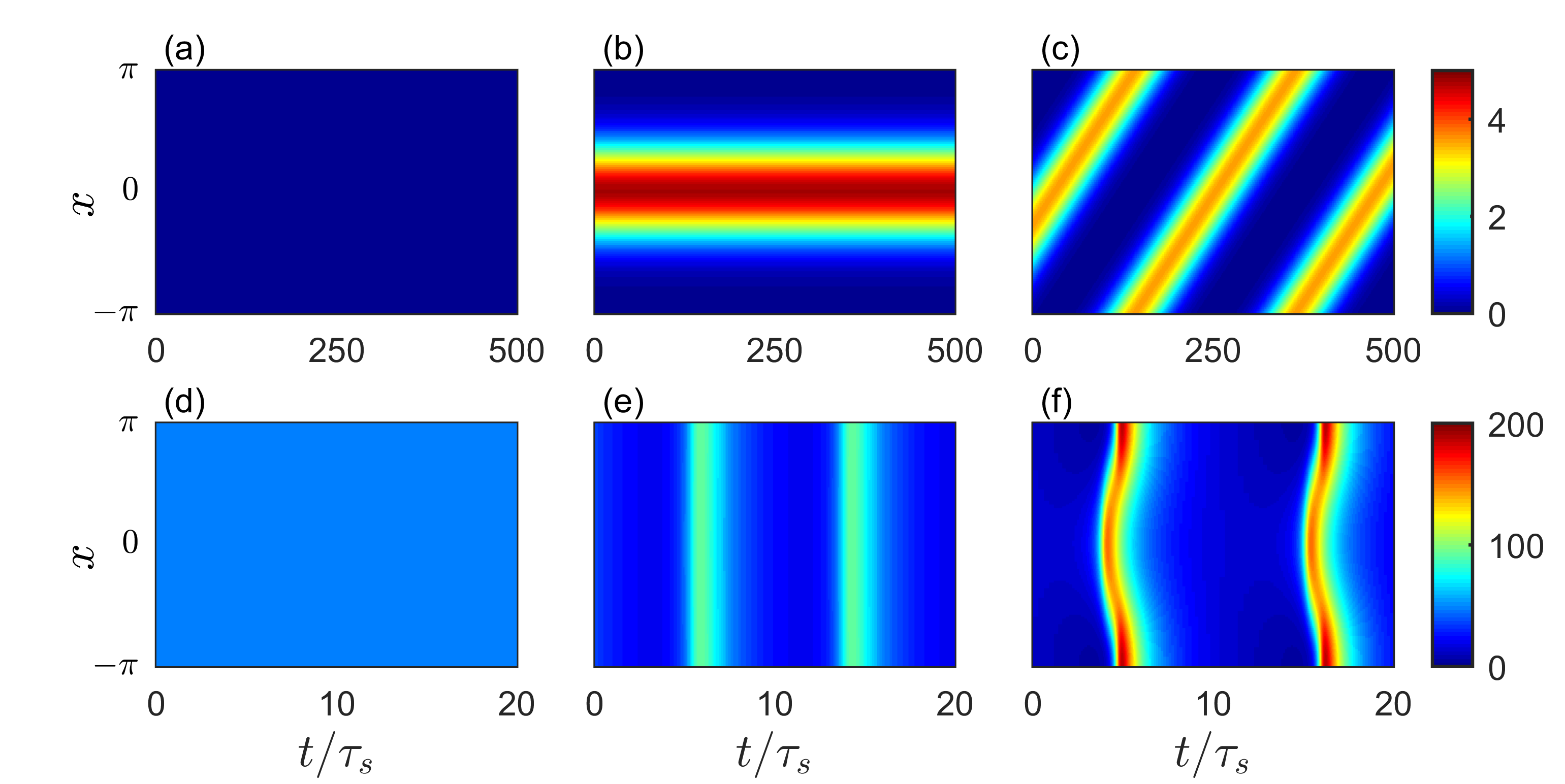}\vspace{-2mm}
\caption{\label{fig:intrin6}
(Color online) Intrinsic Behaviors. The color scale shows $U(x,t)$. 
(a) Silent state. Parameters: $k=0.8$, $\beta=0.2$. 
(b) Static bump. Parameters: $k=0.8$, $\beta=0.005$. 
(c) Moving bump. Parameters: $k=0.8$, $\beta=0.05$. 
(d) Uniform firing. Parameters: $k=1\times10^{-4}$, $\beta=0.02$. 
(e) Homogeneous spikes. Parameters: $k=1\times10^{-4}$, $\beta=0.023$. 
(f) Spikes and anti-spikes. Parameters: $k=1\times10^{-4}$, $\beta=0.0245$. 
For all (a)-(f), $a=0.6$.
}\vspace{2mm}
\includegraphics[width=8.6cm]{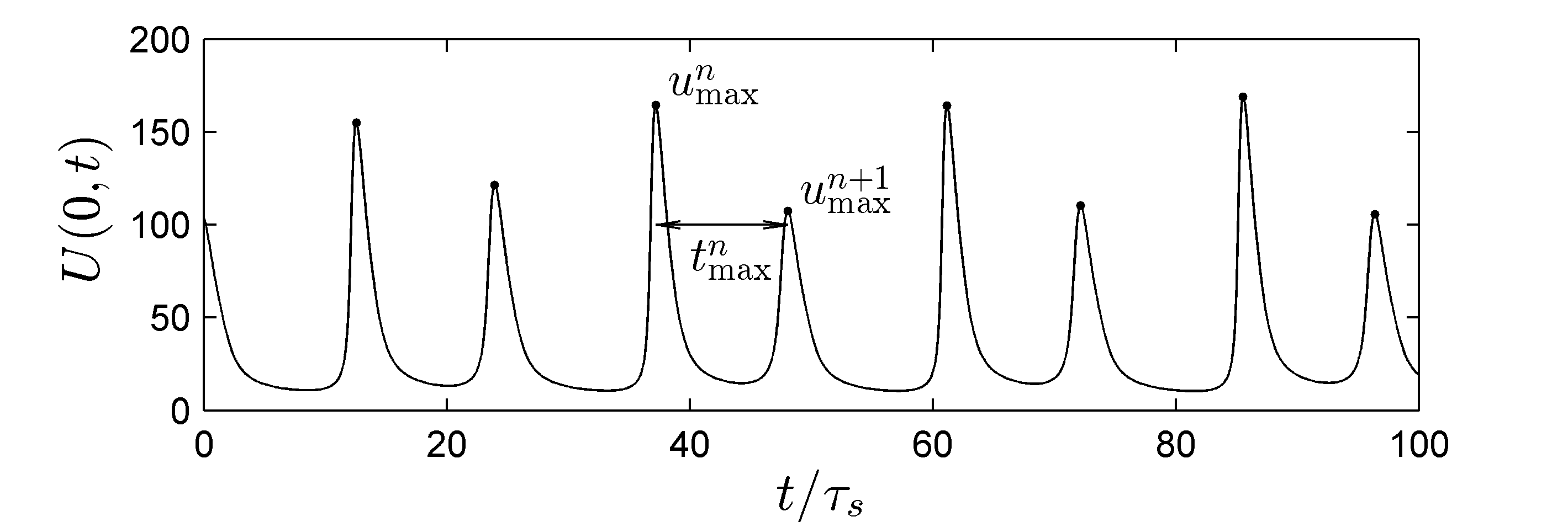}\vspace{-2mm}
\caption{\label{fig:u0_chaotic}
Chaotic Spikes. The trajectory of ${U}(0,t)$.
Parameters: ${k}=3.7\times10^{-4}$, $\beta=0.026999$, and $a=0.6$.
}\vspace{2mm}
\includegraphics[width=8.6cm]{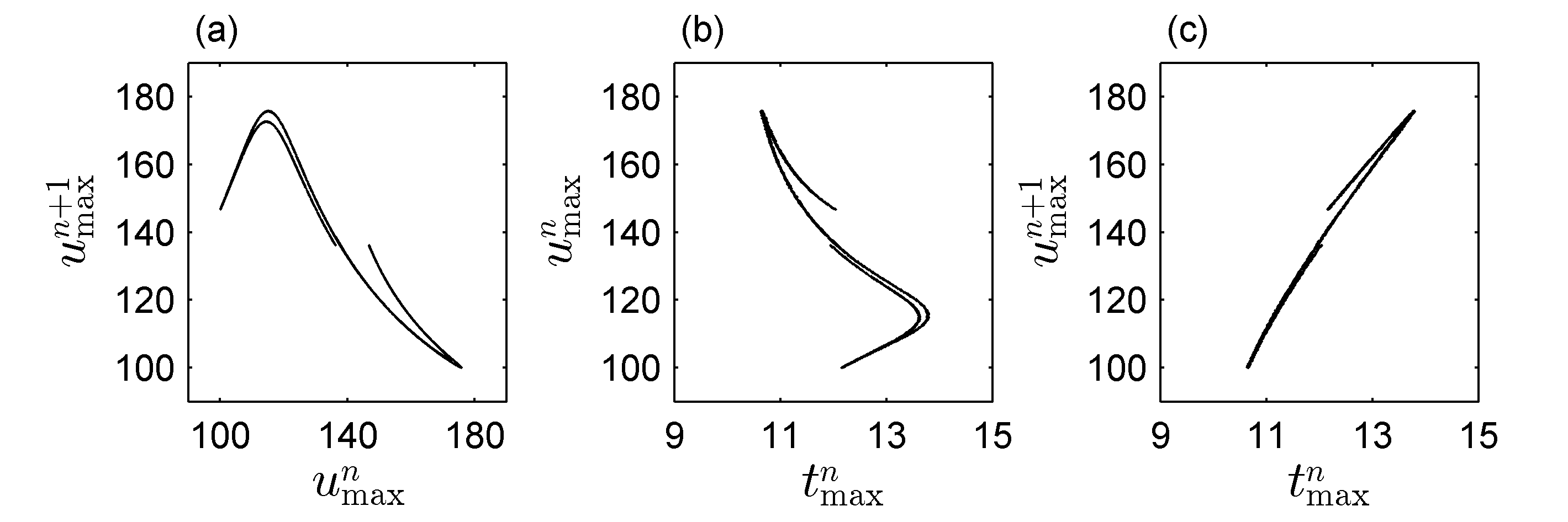}\vspace{-2mm}
\caption{\label{fig:lorenz_maps}
Lorenz maps of chaotic spikes. (a) The relation between 
$u_{\text{max}}^{n+1}$ and $u_{\text{max}}^{n}$. 
(b) The relation between $t_{\text{max}}^{n}$ and 
$u_{\text{max}}^{n}$. (c) The relation between 
$t_{\text{max}}^{n}$ and $u_{\text{max}}^{n+1}$. 
Simulation runs for $50000\tau_{s}$. Parameters are the 
same as in Fig.~\ref{fig:u0_chaotic}.
}
\end{figure}

Various effects of STD on CANN have been studied in the literature.
For instance, Fung \etal\ reported that CANN with STD can 
support four different phases according to strengths of inhibition 
and STD \cite{Fung2012}. In particular, STD can drive traveling 
bumps in the attractor space (see Fig.~\ref{fig:intrin6}(c)). 
These moving profiles happen in the absence of external 
inputs. Additionally, York and van Rossum reported a similar 
result, but with a uniform background current \cite{York2009}.

Moreover, the network's response can be modulated by the interplay
between the STD-driven intrinsic dynamics and a moving stimulus.
Without STD, a bump-shaped neuronal activity profile tracks 
a moving stimulus with a delay.
However, with a proper strength of STD, the network activity profile can move 
ahead of the moving stimulus.
Effectively, the network activity profile is 
located at a future position of the moving stimulus.
It can be used to implement an anticipation mechanism, 
which can compensate inherent delays in the neural system,
thus achieving real-time tracking \cite{NIPS2012_0526}.

\begin{figure}[ht]
\includegraphics[width=8.6cm]{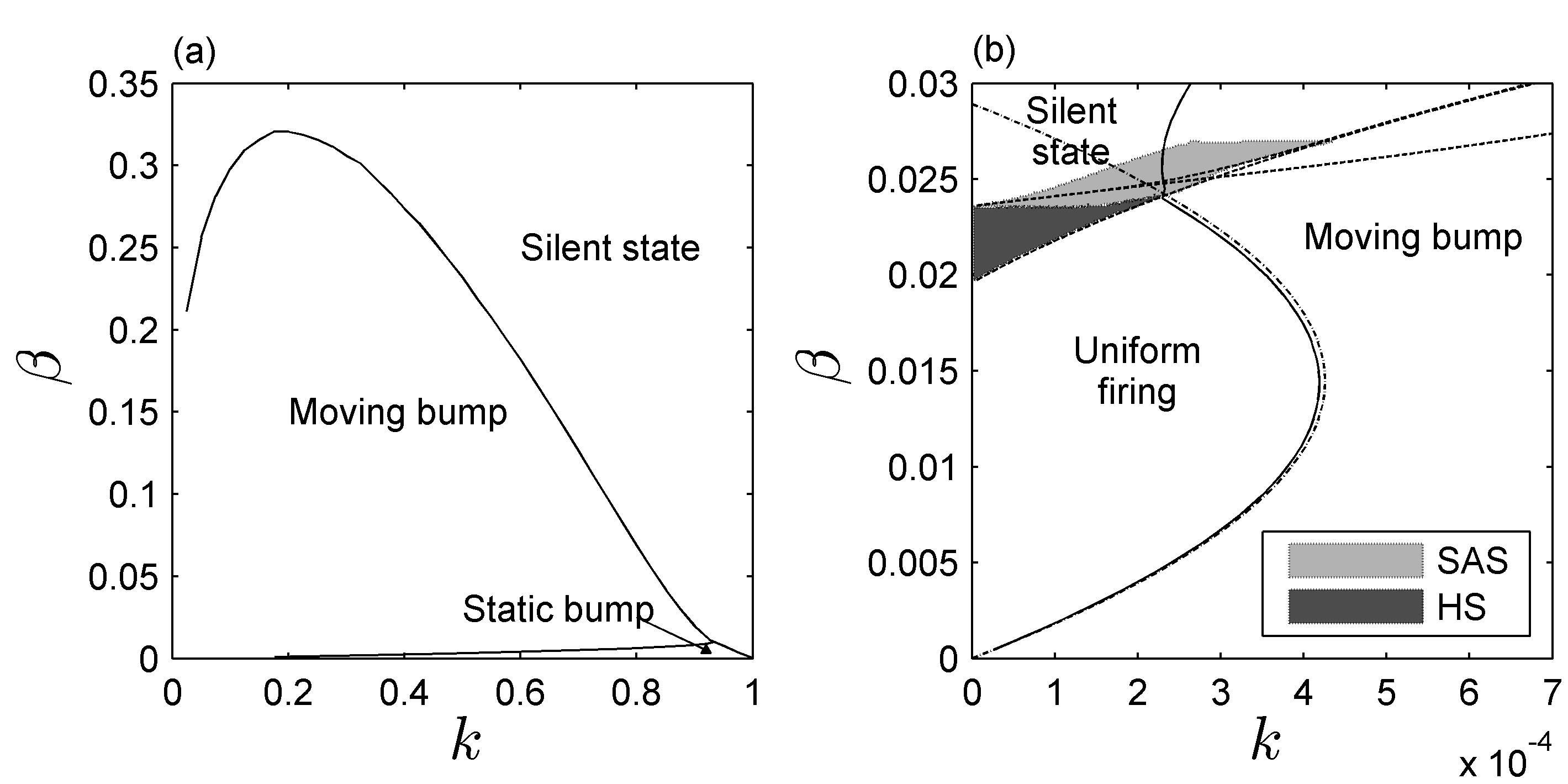}\vspace{-2mm}
\caption{\label{fig:intrin_pd}
Phase diagram of CANN with STD. (a) Moderate global inhibition. 
(b) Weak global inhibition. Moving bump exists on the right 
side of the solid line. Light gray area (SAS) is for spikes 
and anti-spikes. Dark gray area (HS) is for homogeneous spikes. 
Dashed lines are boundaries of the simplified dynamics 
(Eqs.~(\ref{eq:du_uniform}) and (\ref{eq:dp_uniform})). They 
are discussed in region B in Fig.~\ref{fig:simplified_pd}. 
The dot-dashed line is the boundary between the uniform firing 
state and the moving bump state given by Eq.~(\ref{eq:wave_bd_q}).
}
\end{figure}

For a static input, short-term synaptic depression can also drive 
periodic excitements of neuronal activity \cite{Kilpatrick2010}. 
We proposed that these periodic excitements enable CANNs to 
support representations of multiple stimuli almost simultaneously 
\cite{Fung2013}. These periodic excitements provide a plausible 
mechanism for resolution enhancement in transparent motion
\cite{Treue2000, Braddick2002}.

The above examples indicate that in the presence of STD, the
dynamics of CANNs has a rich spectrum. The intrinsic dynamics is 
further enriched in response to external stimuli. So, it is important to 
systematically study the structure of the dynamics of CANNs with STD.
In this paper, we report our 
investigations of CANN with STD in two scenarios:
one with very weak inhibition and 
the other with moderate inhibition in the presence of 
a single static input.

For CANNs with STD and very weak inhibition, the dynamical picture 
is more complicated than that with moderate inhibition strengths, 
which only consists of phases with static bumps, metastatic bumps, 
moving bumps, and the silent state \cite{Fung2012}. 
Depending on parameters, the system can support the following 
behaviors: uniform firing patterns across the whole network, periodic 
excitements of uniform firing, wave instabilities with complex and 
even chaotic firing patterns.

For CANNs with STD and a single static input, the network can 
support various complex firing patterns. Those complex patterns are 
due to the interplay between the STD-driven intrinsic dynamics and 
the external input, depending on strengths of the single static input, 
inhibition and STD. Similar to the previous scenario, some 
complex patterns are chaotic.

The rest of the paper is organized as follows. We will begin with 
an introduction of the model being used throughout the paper. 
Next, we will present the intrinsic dynamics of the CANN with STD 
and very weak inhibition. After that, we will report how the 
dynamics of the system is affected by the external input. Detailed 
behaviors of the network in various phases in the phase diagram will 
also be reported. The relevance of the results will be discussed at the end.

\section{\label{sec:intrinsic}intrinsic behaviors under weak global inhibition}
$N$ neurons are evenly distributed in the space of preferred 
stimuli (in the following simulation results, $N=256$). Neurons are 
labeled by their preferred stimulus $x$. The range of $\{x\}$ is 
$(-L/2,L/2]$. So the size of the space is $L$. Since usually the 
model is applied to the representation of directions or orientations, 
$L=2\pi$ and the periodic boundary condition is imposed. We modify 
the general form of neural field theory and formulate the intrinsic 
dynamics of the neuronal input $\widetilde{U}(x,t)$ 
as \cite{Wilson1972,Amari1977,Fung2010, Fung2012}
\begin{eqnarray}
\tau_s\frac{\partial \widetilde{U}(x,t)}{\partial t}&=
&\rho\int_{-L/2}^{L/2}dx'J(x,x'){p}(x',t)\widetilde{r}(x',t)
\nonumber\\
&&-\widetilde{U}(x,t)+\widetilde{I}(x,t),
\label{eq:u}
\end{eqnarray}
where $\rho$ is the density of neurons in the space of preferred 
stimuli,
$\widetilde{I}(x,t)$ is the external stimulus, 
$J(x,x')$ is the coupling strength between neurons with preferred stimuli
$x$ and $x'$, 
$\widetilde{r}(x',t)$ is the firing rate of neuron $x'$ at time $t$.
They are given by
\begin{eqnarray}
J(x,x')&=&\frac{J_0}{\sqrt{2\pi}a}\exp\left[-\frac{(x-x')^2}{2a^2}\right],
\label{eq:j}\\
\widetilde{r}(x,t)&=&\frac{[\widetilde{U}(x,t)]_+^2}{1+\widetilde{k}\rho
\int dx'[\widetilde{U}(x',t)]_+^2},
\label{eq:r}
\end{eqnarray}
where $\widetilde{k}$ is the strength of the global inhibition and 
$[X]_+\equiv\max(X,0)$. Here we adopt a Gaussian coupling 
and incorporate inhibitory connections into the global inhibition.

${p}(x,t)$ is the availability of neurotransmitters in 
neuron $x$. The dynamics of ${p}(x,t)$ is given by
\begin{equation}
\frac{\partial p(x,t)}{\partial t}=
\frac{1-p(x,t)}{\tau_d}-\widetilde{\beta} p(x,t)\widetilde{r}(x,t),
\label{eq:p}
\end{equation}
where $\tau_d$ is the time scale of neurotransmitter recovery, which 
is chosen to be $\tau_d=50\tau_s$. $\widetilde{\beta}$ is the fraction 
of total neurotransmitters consumed by firing per spike. This implies 
that when the firing rate $\widetilde{r}$ is $0$, $p$ will gradually 
recover to $1$ and the effective connection strength will be exactly 
given by $J(x,x')$. However, $p$ is less than $1$ for active neurons 
and the connection strength is undermined by inefficient transmission. 
For simplicity of analysis, we introduce the rescaled variables and 
parameters: $U\equiv\rho J_0 \widetilde{U}$, 
$\beta\equiv\tau_d \widetilde{\beta}/(\rho J_0)^2$, 
$r\equiv(\rho J_0)^2 \widetilde{r}$,
$k\equiv8\sqrt{2\pi}a\widetilde{k}/(\rho J_0)^2$ and
$I\equiv\rho J_0\widetilde{I}$.
So that we could rewrite Eqs.~(\ref{eq:u})-(\ref{eq:p}) as below.
\begin{eqnarray}
\tau_s\frac{\partial U(x,t)}{\partial t}&=
&\int_{-L/2}^{L/2}dx'\frac{1}{\sqrt{2\pi}a}e^{-\frac{(x-x')^2}{2a^2}}
p(x',t)r(x',t)\nonumber\\
&&-U(x,t)+I(x,t),
\label{eq:ur}\\
\tau_d\frac{\partial p(x,t)}{\partial t}&=&1-p(x,t)-\beta p(x,t)r(x,t),
\label{eq:pr}\\
r(x,t)&=&\frac{[U(x,t)]_+^2}{1+\frac{k}{8\sqrt{2\pi}a}\int dx'[U(x',t)]_+^2}.
\label{eq:rr}
\end{eqnarray}

In the absence of external inputs ($I(x,t)=0$), a variety of interesting 
behaviors have been discovered, such as the static bump, the 
moving bump and the silent state (Fig.~\ref{fig:intrin6}(a)-(c), see also 
\cite{Fung2012}). The static bump state, also known as a persistent 
spatially localized activity state \cite{Camperi1998}, 
is of interest because it can be found in physiological recordings in 
the prefrontal cortex during spatial working memory tasks and other 
systems that encode directional or spatial information, such as head 
direction cells in thalamus and basal ganglia and place cells in the 
hippocampus. The existence and stability of the static bump state 
were first analyzed in detail by Amari \cite{Amari1977}, followed by 
various extensions. Because of the translational invariance of the 
neuronal coupling $J(x,x')$, the center of the static bump can be 
arbitrarily positioned. This is referred to as neutral stability, which 
leads to the naming of continuous attractor neural networks and the 
remarkable tracking ability of this model. 

The moving bump state corresponds to traveling waves which have been 
extensively studied experimentally \cite{Chervin1988, Pinto2005, 
Richardson2005} and theoretically \cite{Golomb1997, Ermentrout2001, 
Pinto2001a, Bressloff2004, Coombes2004}. In our model, 
neurotransmitters are depleted at the bump's position due to the STD. Thus, 
the bump tends to move away to regions where neurotransmitters are 
more available, which is analogous to the spreading of forest fire as the 
fire also moves from places where trees are all burnt out, to places where 
fuels are abundant. This spontaneous movement is proposed to be able 
to compensate various kinds of delays in the neural system and 
therefore facilitate an accurate representation of 
moving stimuli \cite{NIPS2012_0526}. 
However, these behaviors are found where the global inhibition $k$ is 
relatively large. The parameter region where $k$ is very small is not 
much explored yet.


\begin{figure}[t]
\includegraphics[width=8.6cm]{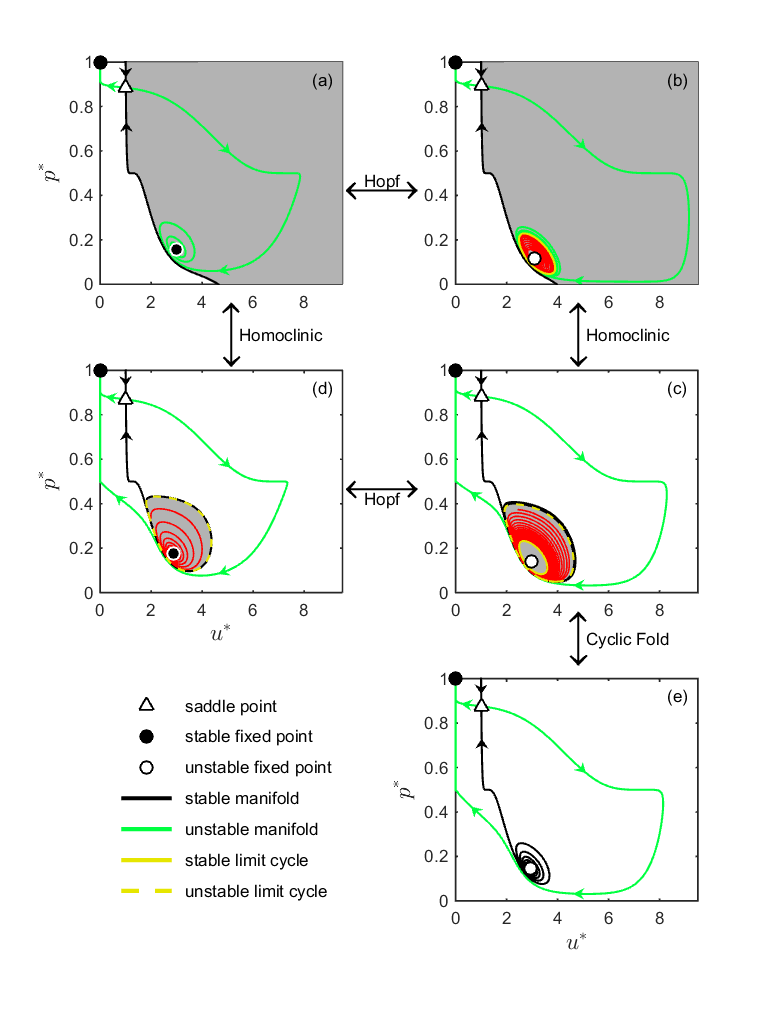}
\caption{\label{fig:limit_cycle}
(Color online) Different behaviors of simplified homogeneous dynamics. 
Variables are rescaled as $u^*=u^{0.3}$ and $p^*=(p - 0.5)^5 /0.5^4 + 0.5$ 
to make a readable presentation of trajectories while keeping topological 
equivalence. Gray area is the basin of attraction for the stable fixed 
point or the stable limit cycle. Some trajectories are colored red to 
illustrate their approach to a stable limit cycle or stable fixed point. 
(a) Uniform firing without 
an unstable limit cycle. Parameters: $ k = 5 \times 10^{-4}$, $ \beta 
= 0.025$. (b) Homogeneous spikes without an unstable limit cycle. 
Parameters: $ k = 1 \times 10^{-4}$, $ \beta = 0.023$. (c) Homogeneous 
spikes with an unstable limit cycle. Parameters: $ k = 2.9 \times 10^{-4}$, 
$ \beta = 0.0253$. (d) Uniform firing with an unstable limit cycle. 
Parameters: $ k = 7 \times 10^{-4}$, $ \beta = 0.0285$. (e) Silent 
state. Parameters: $ k = 3 \times 10^{-4}$, $ \beta = 0.027$. For 
(a)-(e), $a=0.6$.}
\end{figure}

\subsection{Phase Diagram}
The global inhibition plays an important role in shaping the bump states. 
Regions with low activity will be suppressed by regions with high 
activity through the global inhibition, thus the neuronal activity will 
be localized to form a state with only one bump, either moving or 
static. In the very weak global inhibition scenario, we expect that 
neural activity would be more uniform and synchronized 
\cite{Hansel1995}. In the uniform firing state, the firing rates of all 
neurons are uniform and time-independent (Fig.~\ref{fig:intrin6}(d)).

Another possible effect of very weak global inhibition is that, neuronal 
activity will be relatively higher and more neurotransmitters are 
consumed, which induces population spikes. The neuronal activity grows up 
very fast due to the weak inhibition and a large amount of neurotransmitters 
are consumed. Then, the activity will die down because of the STD. This is 
called a population spike (also called ensemble synchronization) \cite{Loebel2002, Mark2012}. After the 
activity dies down, neurotransmitters will gradually recover to the 
level that can support another population spike. These population spikes 
show various dynamics, including homogeneous spikes and 
the spike-and-anti-spike state, in the very weak global inhibition 
scenario (Fig.~\ref{fig:intrin6}(e-f)).

In the homogeneous spike state, all neurons in the network are 
synchronized in population spikes (Fig.~\ref{fig:intrin6}(e)). In the 
spike-and-anti-spike state, one spike emerges at a certain place in the 
network, then splits into two symmetric branches of moving bumps, which 
will collide with each other at the opposite side of the network and form 
the anti-spike, and the firing activities stop. The cycle then repeats itself 
periodically (Fig.~\ref{fig:intrin6}(f)). 

In some parameter regions, spikes and anti-spikes can be chaotic. In 
the simulation, the initial condition is chosen to make spikes and 
anti-spikes appear at $x=0$ and $x=\pi$, respectively. We could see 
the chaotic behavior by just looking at the dynamics at $x=0$ 
(Fig.~\ref{fig:u0_chaotic}). Define $u_{\text{max}}^{n}$ as the 
$n^{\rm th}$ peak of ${U}(0,t)$ and $t_{\text{max}}^{n}$ as the temporal 
interval between the $n^{\rm th}$ and $(n+1)^{\rm th}$ peaks of ${U}(0,t)$.
By examining the relation between $u_{\text{max}}^{n+1}$ and 
$u_{\text{max}}^{n}$ (Fig.~\ref{fig:lorenz_maps}(a)), we could see it 
shows chaotic features. The relation between temporal intervals between 
spikes and the height of spikes (Fig.~\ref{fig:lorenz_maps}(b) and (c)) 
shows that the height of a spike is almost linearly dependent on the 
resting time before that spike, rather than after that spike. Since 
longer resting times mean fuller recovery of neurotransmitters, the 
population spike following a long resting time is able to reach a 
greater height.

Simulations are performed to find the boundaries of these behaviors in 
the phase space (Fig.~\ref{fig:intrin_pd}). In the phase diagram, uniform 
firing phase is found where ${k}$ is less than $4\times10^{-4}$ and 
${\beta}$ is less than $0.02$. When ${k}$ increases, the uniform firing state
becomes moving bumps. This is because stronger global inhibition will 
suppress the homogeneity of the dynamics. When ${\beta}$ increases, 
the uniform firing state becomes homogeneous spikes (dark gray area in 
Fig.~\ref{fig:intrin_pd}(b)). The homogeneity of the dynamics is 
maintained, but stronger STD introduces temporal modulation. Spikes 
and anti-spikes (light gray area in Fig.~\ref{fig:intrin_pd}(b)) are 
found when ${\beta}$ is even larger. Some part of this area overlaps 
with moving bump region, which indicates the coexistence of the two behaviors.
\begin{figure}[t]
\includegraphics[width=6.9cm]{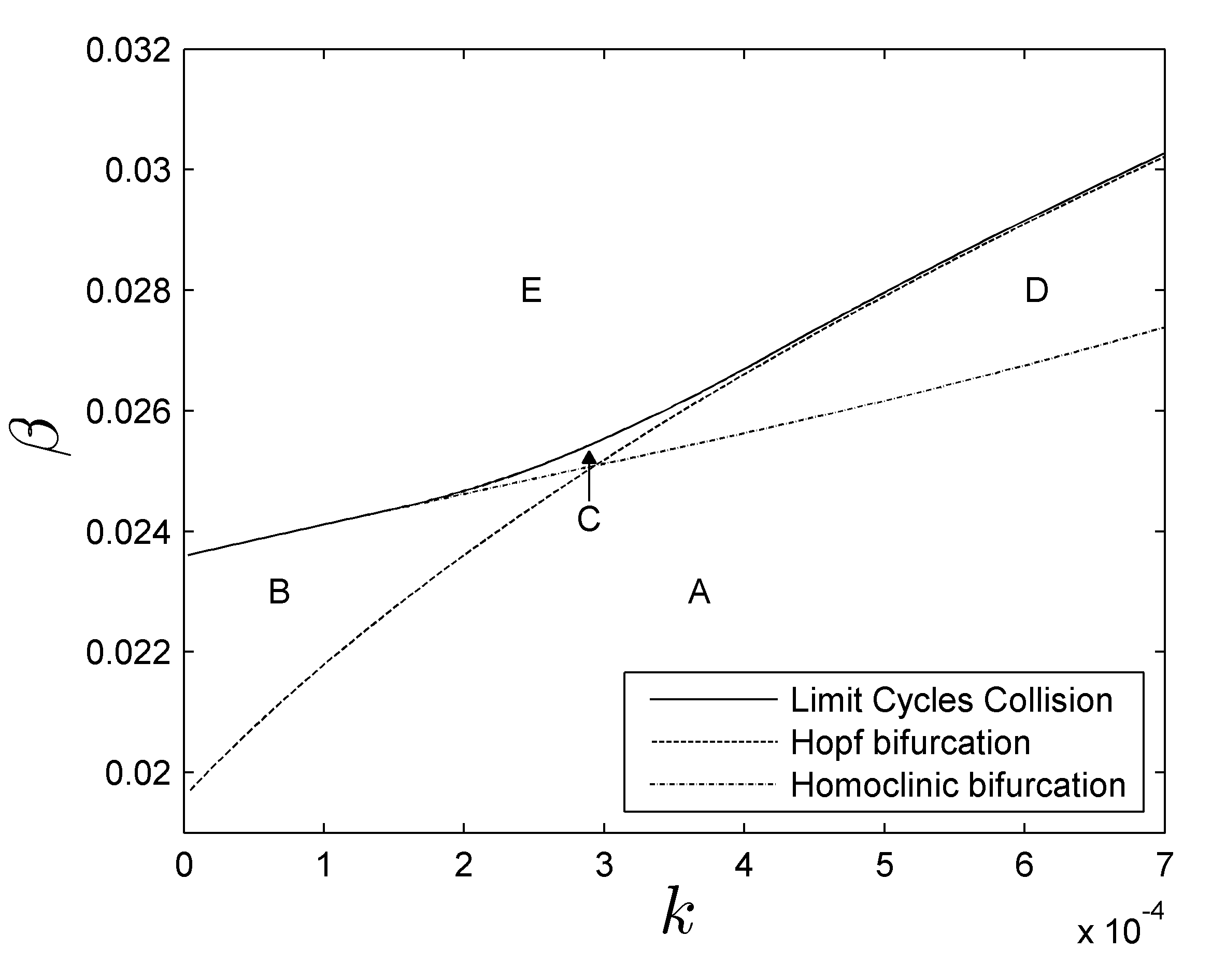}
\caption{
\label{fig:simplified_pd}
Phase diagram of simplified homogeneous dynamics for interaction 
range $a = 0.6$.  Region A, B, C, D and E correspond to behaviors in 
Fig. \ref{fig:limit_cycle}(a), (b), (c), (d) and (e), respectively.}
\end{figure}

\subsection{Simplified homogeneous dynamics}
In the phase diagram shown in Fig.~\ref{fig:intrin_pd}(b), there is a phase 
called uniform firing.  To study the stability of this pattern, we simplified 
the dynamics in Eqs. (\ref{eq:ur}) and (\ref{eq:pr}) to be a pair of spatially 
independent differential equations:
\begin{eqnarray}
\tau_s \frac{\partial u\left(t\right)}{\partial t} 
&=& -u\left(t\right) + \frac{p\left(t\right)u\left(t\right)^2}{B} J_a, \label{eq:du_uniform} \\
\tau_d \frac{\partial p\left(t\right)}{\partial t} 
&=& 1-p\left(t\right) - \frac{{\beta} p\left(t\right)u\left(t\right)^2}{B}.\label{eq:dp_uniform}
\end{eqnarray}
Here, $u(t)$ represents the spatially independent $U(x,t)$. $J_a\equiv 
{\rm erf} [L/(\sqrt{8}a)]$ and $B\equiv 1+ k L / (8\sqrt{2\pi}a)u^2$.  
The nullclines for $du/dt$ and $dp/dt$ are 
\begin{eqnarray}
u&=&\frac{pu^2}{B}J_a,\nonumber\\
1-p&=&\frac{{\beta}pu^2}{B}.\label{eq:homo_nul}
\end{eqnarray}
Intersections between nullclines are fixed point solutions. At the 
intersections,
\begin{equation}
u=
\frac{J_a\pm\sqrt{J_a^2-
4\left(\beta+\frac{ kL}{8\sqrt{2\pi}a}\right)}}
{2\left(\beta+\frac{ kL}{8\sqrt{2\pi}a}\right)}.
\label{eq:u_fixed_uniform}
\end{equation}

Besides fixed point solutions, limit cycles could also appear through the Hopf 
bifurcation or the homoclinic bifurcation \cite{EABT3}.
In Fig.~\ref{fig:limit_cycle}, there are 
typical phase portraits in the simplified system for different values of $k$ 
and $\beta$. In Fig.~\ref{fig:limit_cycle}(a), there are two stable fixed 
points, one representing the silent state and the other representing the 
uniform firing state. The stable manifold of the saddle point is the 
boundary separating basins of attraction of the two stable fixed points. 
Temporal oscillations exist in the transient state, but the firing rate is
time-independent at the steady state. The fixed point for uniform firing 
loses its stability in Fig.~\ref{fig:limit_cycle}(b), and a stable limit cycle 
appears around it, which demonstrates a Hopf bifurcation. The stable 
limit cycle corresponds to homogeneous spikes. Between 
Fig.~\ref{fig:limit_cycle}(a) and (d), a homoclinic bifurcation occurs. The 
unstable manifold of the saddle point touches its stable manifold, resulting 
in a homoclinic orbit, which is also an unstable limit cycle encircling the 
basin of attraction of the uniform firing fixed point. Beyond the 
homoclinic bifurcation, the unstable limit cycle is no longer homoclinic 
and shrinks in Fig.~\ref{fig:limit_cycle}(d). Although this unstable limit 
cycle does not affect the stability of the fixed points, it confines the basin 
of attraction of the uniform firing state. Between Fig.~\ref{fig:limit_cycle}(b)
and (c), the same homoclinic bifurcation happens and the unstable limit 
cycle is the basin boundary of the stable limit cycle. Between 
Fig.~\ref{fig:limit_cycle}(c) and (d), the Hopf bifurcation same as the one 
between Fig.~\ref{fig:limit_cycle}(a) and (b) happens. Between 
Fig.~\ref{fig:limit_cycle}(c) and (e), the stable limit cycle and the unstable 
limit cycle approach each other and collide, which makes a fold bifurcation
of cycles \cite{EABT3}.

To study the stability of the above fixed point solutions, we assume  that
$u(t) = u_0 + u_1(t)$ and $p(t) = p_0 + p_1(t)$, where $u_0$ and $p_0$ 
are the fixed point solution being investigated.  By linearizing Eqs. 
(\ref{eq:du_uniform}) and (\ref{eq:dp_uniform}), we have

\begin{equation}
\frac{d}{dt}\left(\begin{array}{c}
u_{1}\\
p_{1}
\end{array}\right)
=\left(\begin{array}{cc}
\frac{2p_0 u_0 J_a - B}{\tau_s B} & 
\frac{u_0{}^2 J_a}{\tau_s B}\\
-\frac{2{\beta}p_0u_0}{\tau_d B} & 
-\frac{B +  \beta u_0{}^2}{\tau_d B}
\end{array}\right)
\left(\begin{array}{c}
u_{1}\\
p_{1}
\end{array}\right).
\label{eq:uniform_linearization}
\end{equation}

There are two non-zero fixed point solutions to Eqs.~(\ref{eq:du_uniform}) and 
(\ref{eq:dp_uniform}): one with large $u$ and the other with smaller $u$, as 
shown in Eq. (\ref{eq:u_fixed_uniform}) (see Appendix \ref{sct:apd_a}). 
At the fixed point solution with smaller $u$, the determinant of the matrix in 
Eq.~(\ref{eq:uniform_linearization}) is negative. Hence it is a saddle point 
(triangles in Fig. \ref{fig:limit_cycle}).

At the fixed point solution with larger $u$, the determinant is always 
larger than zero.  The stability condition is given by the trace of the 
matrix.  In order to have a stable fixed point solution, the trace should be 
less than zero.  After some algebra, the phase boundary of uniform firing 
is given by
\begin{equation}
\begin{cases}
{{k}}=\frac{8\sqrt{2\pi}aJ_{a}^{2}}{L}\left(\frac{\tau_{s}}{\tau_{d}}\right)^{2}
\frac{B-1}{\left(2-B\right)^{2}}\\ 
{{\beta}}=\frac{\tau_{s}J_{a}^{2}}{\tau_{d}}
\frac{2-\left(1+{\tau_{s}}/{\tau_{d}}\right)B}{\left(2-B\right)^{2}}
\end{cases},\label{eq:homo_hopf}
\end{equation}
where $B>1$. This is a parametric expression for the boundary of the Hopf 
bifurcation (the dashed line is Fig.~\ref{fig:simplified_pd}). Other 
bifurcation curves in the phase diagram of this simplified system
(Fig.~\ref{fig:simplified_pd}) can be 
found using the numerical continuation package MATCONT \cite{Matcont2003}.  
We can see 
that this simplified model captures homogeneous behaviors very well.  
However, for inhomogeneous behaviors, we need to consider the wave 
stability of the fixed point solutions of Eqs. (\ref{eq:ur}) and 
(\ref{eq:pr}).

\subsection{Wave stability of the uniform firing}

To understand why moving bump disappears in such low values of 
inhibition and what small parameter characterizes the existence of 
uniform firing, we consider fluctuations with wave vector $q$,
so that $U(x,t)=u_{0}+u_{1}(t)e^{iqx}$ 
and $p(x,t)=p_{0}+p_{1}(t)e^{iqx}$, where $u_{0}$ and $p_{0}$ are 
given by the steady state solution.
Putting these into Eqs.~(\ref{eq:ur}) and (\ref{eq:pr}), 
and keeping terms up to the first order, we have
\begin{equation}
\frac{d}{dt}\left(\begin{array}{c}
u_{1}\\
p_{1}
\end{array}\right)
=\left(\begin{array}{cc}
\frac{2p_0 u_0 Q - B}{\tau_s B} & 
\frac{u_0{}^2 Q}{\tau_s B}\\
-\frac{2{\beta}p_0u_0}{\tau_d B} & 
-\frac{B +  \beta u_0{}^2}{\tau_d B}
\end{array}\right)
\left(\begin{array}{c}
u_{1}\\
p_{1}
\end{array}\right),
\label{eq:wave_linearization}
\end{equation}
where $Q=\int_{-L/2}^{L/2}dx'\frac{1}{\sqrt{2\pi}a}
\exp[-\frac{\left(x-x'\right)^{2}}{2a^{2}}-iq\left(x-x'\right)]$. 
Note that here we just consider wave 
fluctuations, thus $B$ is not affected by the fluctuations to the first order.

To evaluate  $Q$, we can express the  integral in terms of 
Hermite polynomials,
%
\begin{eqnarray}
Q&=&e^{-a^{2}q^{2}/2}\bigg[\text{erf}\left(\frac{L}{\sqrt{8}a}\right)\nonumber\\
&&-2\sum_{n=1}^{\infty}\frac{e^{-L^{2}/8a^{2}}}{\sqrt{\pi}n!}\left(-\frac{iqa}{\sqrt{2}}\right)^{n}H_{n-1}\left(\frac{L}{\sqrt{8}a}\right)\bigg].
\label{eq:wave_u_integral_3}
\end{eqnarray}

Owing to the smallness of the Gaussian factor in the higher order terms, we 
can approximate $Q$ by $J_{a}e^{-a^{2}q^{2}/2}$, where 
$J_{a}\equiv\text{erf}\left[{L}/{(\sqrt{8}a)}\right]$, or approximately $1$ 
when $a \ll L$. 

For fixed point solutions, the stability condition is the trace of the 
stability matrix, $T\le0$. Hence the phase boundary is given by 
$T=2p_0u_0J_{a}e^{-a^{2}q^{2}/2}/B-1-\left(1+{{\beta}u_0{}^{2}}/{B}\right){\tau_{s}}/{\tau_{d}}=0$. 
Combining with the steady state solutions, we can express the stability 
condition as
\begin{equation}
2e^{-a^{2}q^{2}/2}-1=\frac{\tau_{s}}{\tau_{d}}\left(1+\frac{{\beta}u_0{}^{2}}{B}\right).
\label{eq:wave_bd_q}
\end{equation}

Numerical solutions show that the instability comes from the long 
wavelength mode. For a ring model of length $L$, the fundamental mode has 
a wave number $q=2\pi/L$. Hence for $L=2\pi$, $q=1$.
Combining with results on the steady state solutions, 
the boundary of wave stability is 
\begin{equation}
{k}=\frac{8\sqrt{2\pi}a}{L}\left(\frac{p_0}{1-p_0}{\beta}
-\frac{1}{\left(1-p_0\right)^{2}}{\beta}^{2}\right),
\label{eq:wave_bd_kb}
\end{equation}
where the value of $p_0$ on the boundary is 
$(2e^{-2\pi^2 a^2 / L^2}-1)^{-1}{\tau_{s}}/{\tau_{d}}$. This boundary is 
plotted as the dot-dashed line in Fig.~\ref{fig:intrin_pd}(b) and it agrees 
very well with the boundary between the uniform firing and the moving bump. The 
maximum of $k$ on this boundary is
\begin{equation}
k_{\rm{max}}=
\frac{2\sqrt{2\pi} J_{a}^2}
{\left(2e^{-2\pi^2 a^2 / L^2}-1\right)^{2}}
\cdot
\frac{{\tau_{s}}^2 a}{{\tau_{d}}^2 L}
.\label{eq:k_max}
\end{equation}
This shows that the smallness of $k$ comes from the scaling
$\left(\tau_{s}/\tau_{d}\right)^{2}a/L$.
On the other hand, the maximum of $\beta$ on the boundary is
\begin{equation}
{\beta}_{\rm{max}}=
\frac{2e^{-2\pi^2 a^2 / L^2}-1-\tau_{s}/\tau_{d}}
{\left(2e^{-2\pi^2 a^2 / L^2}-1\right)^2}
\cdot
\frac{J_{a}^{2}\tau_{s}}{\tau_{d}}
.\label{eq:b_max}
\end{equation}
This shows that the smallness of $\beta$ comes from 
the scaling $\tau_{s}/\tau_{d}$.

\section{\label{sec:static}responses to a single static input}
In this section, we will discuss the network behavior in the presence of
a single static input and moderate global inhibition. For the external input in 
Eq.~(\ref{eq:ur}), we adopt the form 
$I(x,t)=A\exp [-{(x-z)^{2}}/({2a_A^{2}})]$, where $z$ is the center of that 
input (without loss of generality, $z=0$ in this work), $A$ is the 
strength of the input, and $a_A$ is the width of the input.
Note that the behavior of this system is controlled by 
three parameters, namely the strength of the global inhibition $k$,
the strength of the STD $\beta$ 
and the strength of the external input $A$. 
In this work, different response patterns are discussed in the parameter 
space spanned by these three quantities.

\begin{figure}[t]
\includegraphics[width=8.6cm]{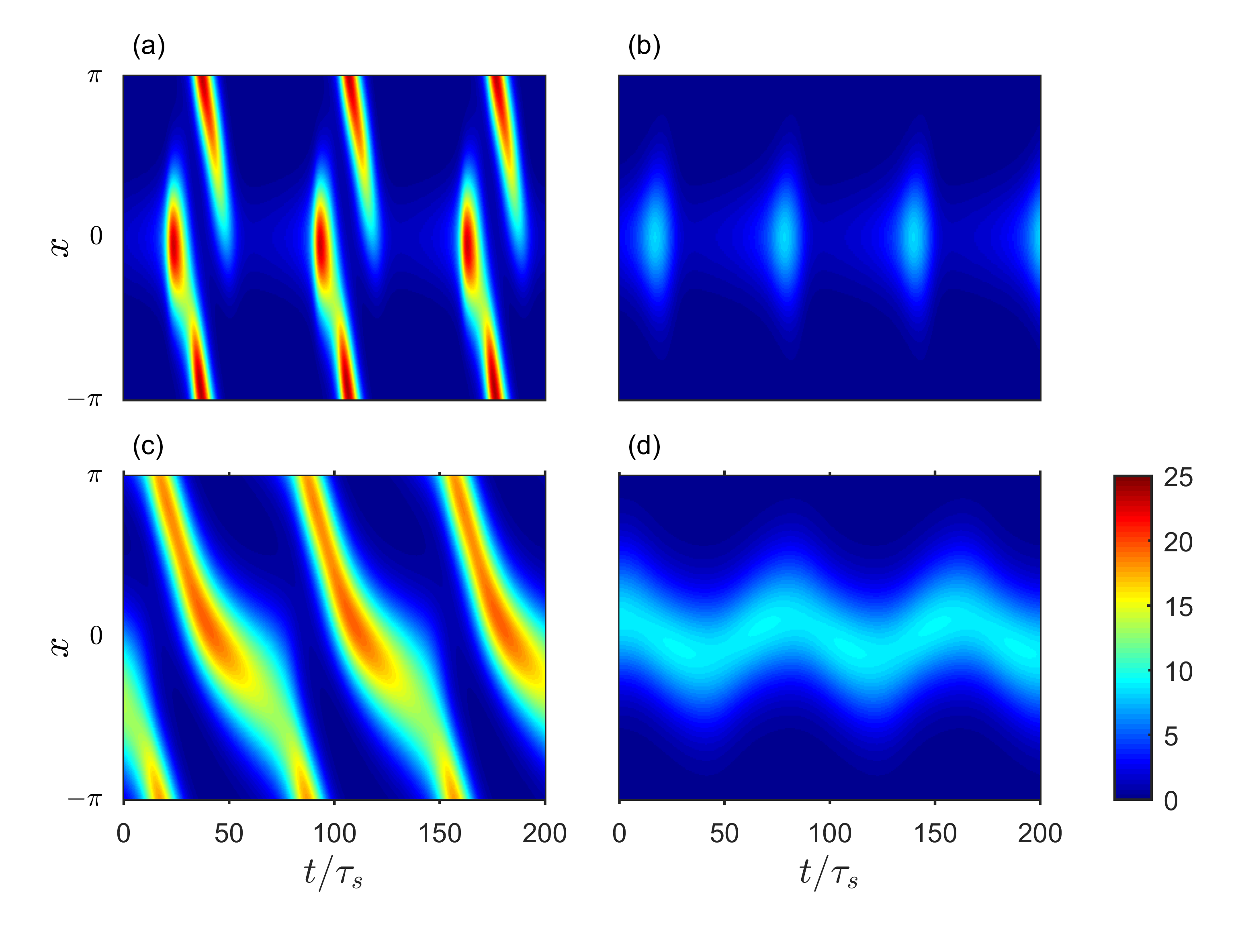}
\caption{\label{fig:4basic}
(Color online) Four basic dynamic responses to a single static input 
in CANN with STD. 
The color scale shows the firing rate $r(x,t)$. (a) Emitter. Parameters: 
$k=0.2$, $\beta=0.3$. (b) Population spikes. Parameters: $k=0.3$, 
$\beta=0.4$. (c) Moving bump. Parameters: $k=0.3$, $\beta=0.1$. (d) 
Slosher. Parameters: $k=0.5$, $\beta=0.1$. For all (a)-(d),
$A=0.8$, $a=a_A=48^{\circ}=0.8378$.
}
\end{figure}
\begin{figure}[t]
\includegraphics[width=8.6cm]{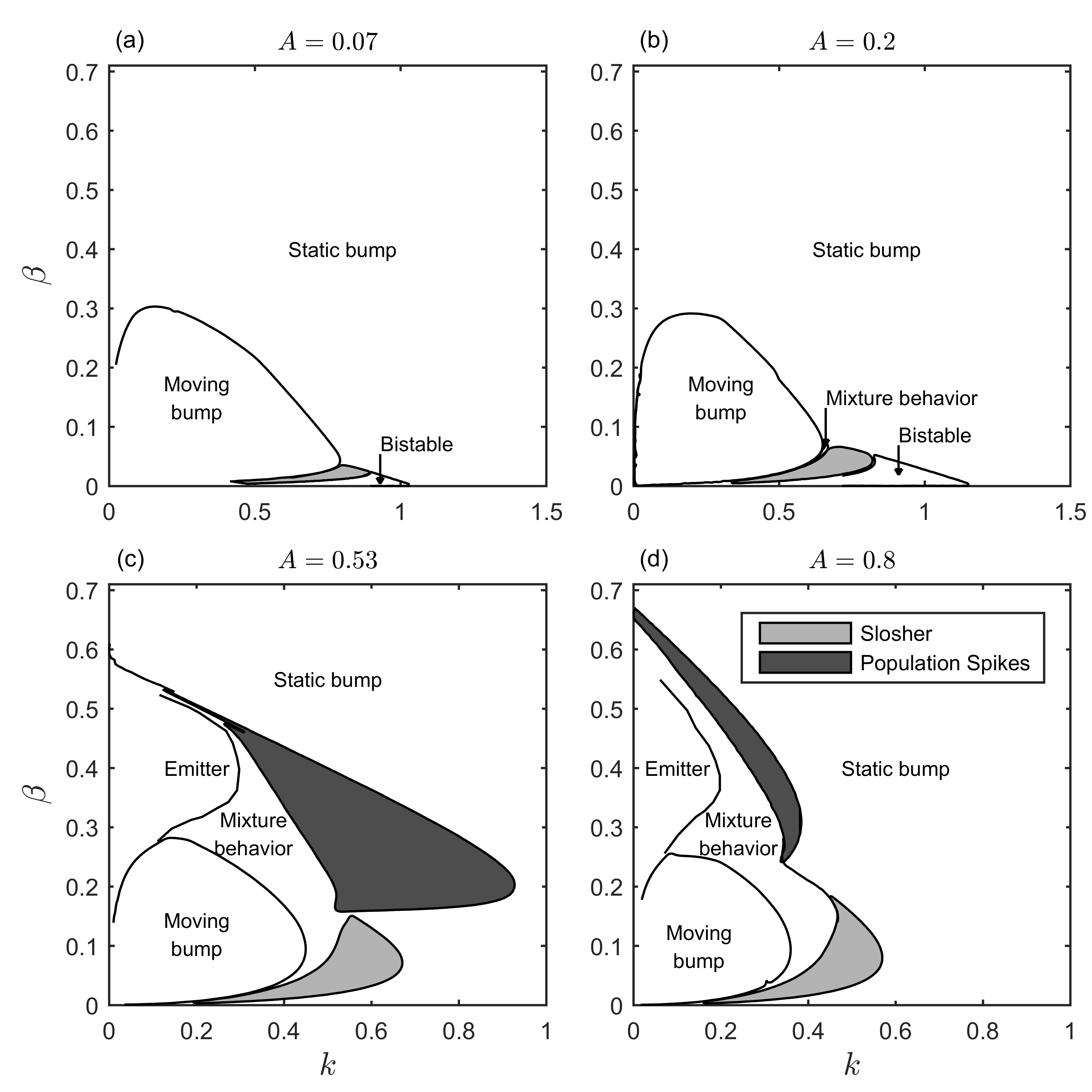}
\caption{\label{fig:PD_As}
Phase diagrams for the four basic responses in the space of $k$ and $\beta$ 
with different values of $A$. $a=0.5$ and $a_A=\sqrt{2}/2$.
}
\end{figure}

\subsection{Four Basic Dynamic Response Patterns}

The static bump is expected to be the simplest form of response. However, 
in a very large region of the parameter space, static bumps are unstable and 
much more interesting response patterns emerge. Among them, there are four 
basic dynamic patterns through which we can understand 
the general property of this system (Fig.~\ref{fig:4basic}(a)-(d)).

One response pattern is the moving bump. Moving bumps result from the 
mobility of the neural field enhanced by the STD. Once a bump 
is built, neurotransmitters are depleted in the bump region, leading to a 
tendency of the bump to move away to fresher regions. As an intrinsic 
behavior, the moving bump will keep its profile and its speed all the time 
(Fig.~\ref{fig:intrin6}(c)). However, while the static input is imposed, the 
speed and profile of the bump will change when the bump crosses the input. 
The bump is higher and faster when approaching to the input, while weaker and 
slower when leaving the input, because the external input tends to attract 
the bump (Fig.~\ref{fig:4basic}(c)).

When the attraction provided by the external input is strong and the mobility 
enhanced by the STD is not sufficient for the bump to overcome 
the attraction of the input, 
the bump gets trapped and moves side-to-side around the external input.
It is called a slosher (Fig.~\ref{fig:4basic}(d)) \cite{Folias2011}.

\begin{figure}[t]
\includegraphics[width=8.6cm]{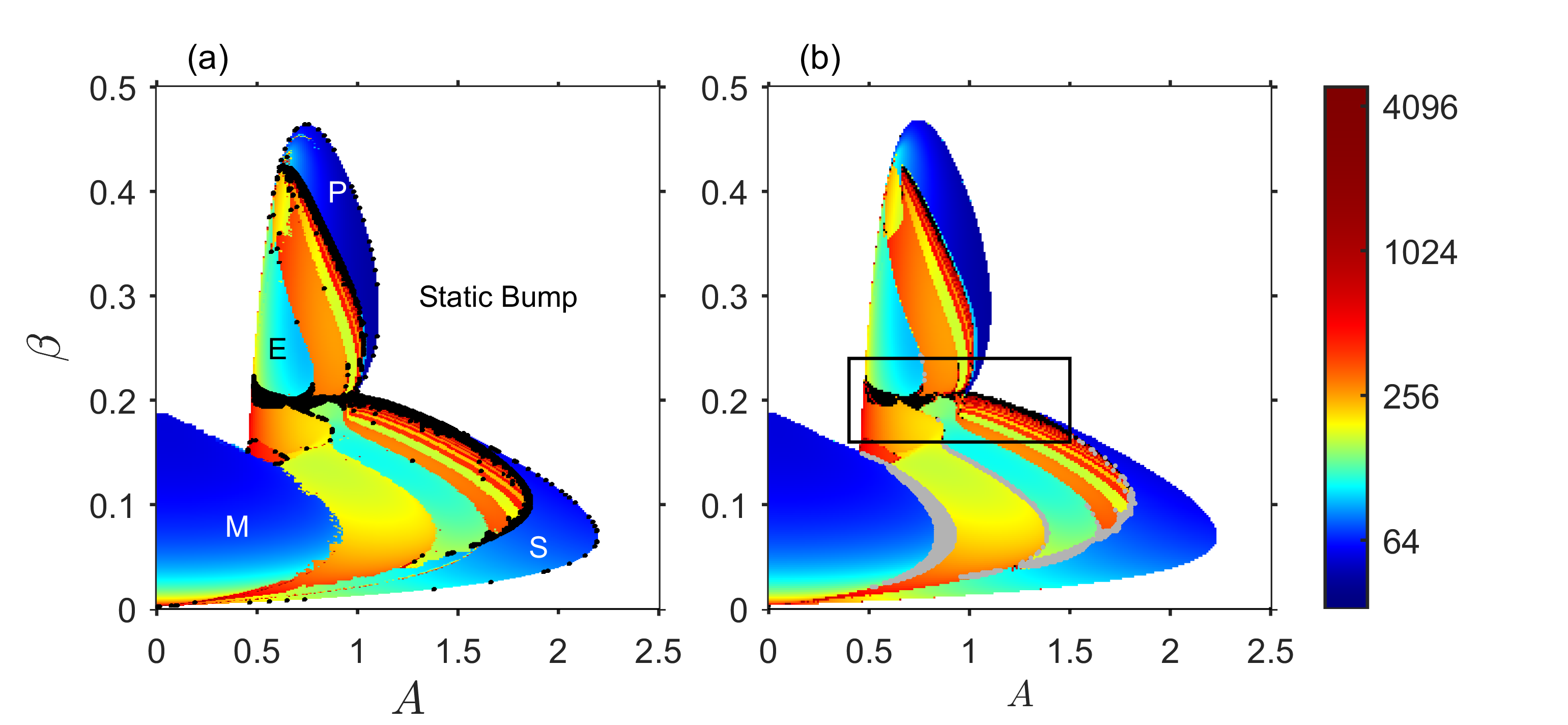}\vspace{-2mm}
\caption{\label{fig:pdk03_compare}
(Color online) Phase diagram in the space of $A$ and $\beta$. The color 
code indicates lengths of the periods in log scale. Black dots means the 
period of the response is too long to be detected by the program, suggesting 
the response is possibly chaotic. The global inhibition strength $k=0.3$. 
$a=a_A=0.8378$.
(a) Full model simulation. M, E, P and S represent ``moving bump'', 
``emitter'', ``population spikes'' and ``slosher'', respectively. (b) 
Numerical solutions of the second order Fourier series expansion. The 
gray regions are bistable regions where the lengths of the periods can be 
either of two different values. 
The box in (b) encircles the region where Lyapunov 
exponents are computed and shown in Fig.~\ref{fig:lle}.
}
\vspace{2mm}
\includegraphics[width=8.6cm]{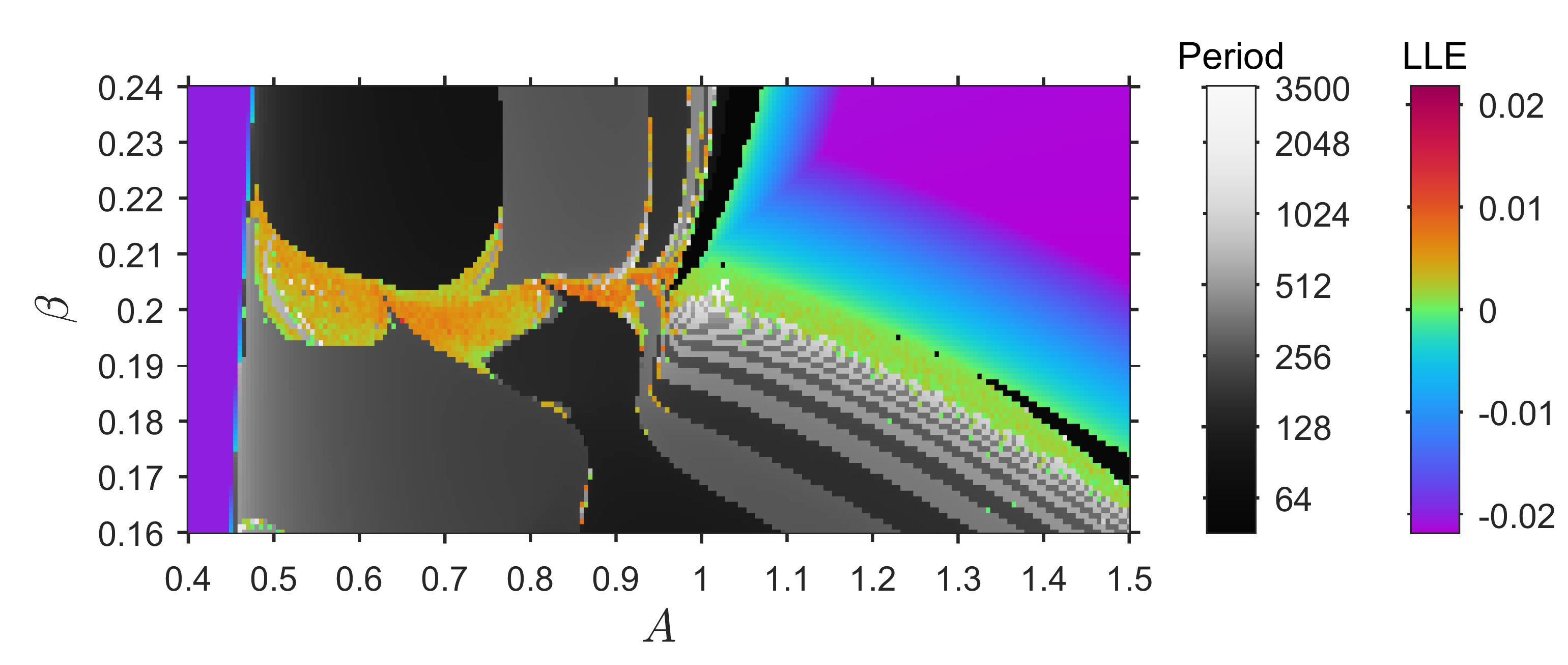}\vspace{-2mm}
\caption{\label{fig:lle}
(Color online) Largest Lyapunov exponents (LLE) computed on the equations of 
second order Fourier series expansion. Parameters are the same as those in 
Fig.~\ref{fig:pdk03_compare}. Gray scale denotes the lengths of the periods for 
periodic behaviors.
}
\end{figure}
\begin{figure}[t]
\includegraphics[width=8.6cm]{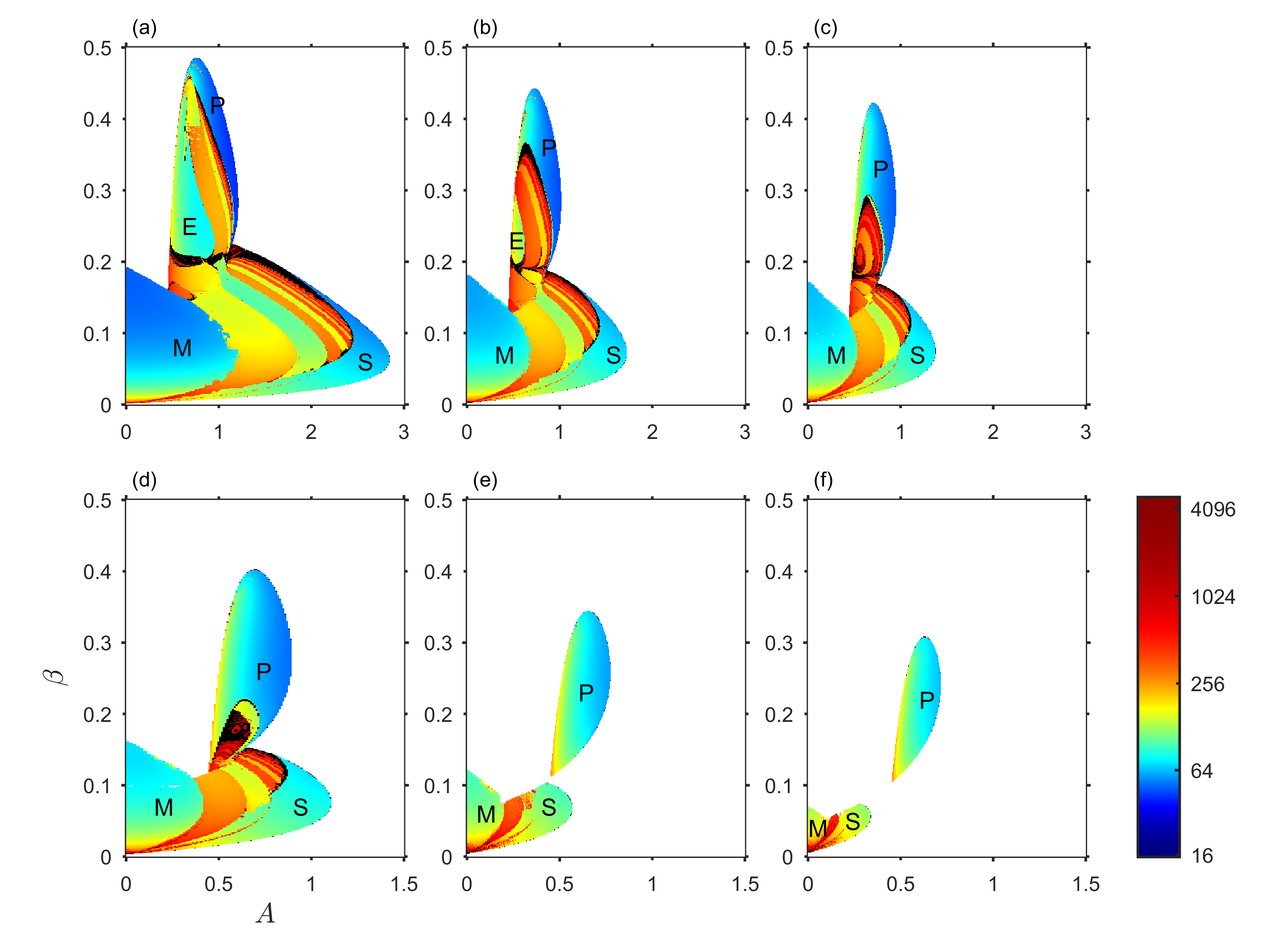}\vspace{-2mm}
\caption{\label{fig:pd_kkk}
(Color online) Phase diagrams in the space of $A$ and $\beta$ with different 
values of $k$. $k$ is $0.25$ in (a), $0.35$ in (b), $0.40$ in (c), $0.45$ 
in (d), $0.60$ in (e) and $0.70$ in (f). The color code indicates
the lengths of the periods in log scale. 
Black dots mean that the period of the response is too long 
to be detected by the program, suggesting the response is possibly 
chaotic. M, E, P and S represent ``moving bump'', ``emitter'', 
``population spikes'' and ``slosher'', respectively.
$a=a_A=0.8378$.
}
\includegraphics[width=8.6cm]{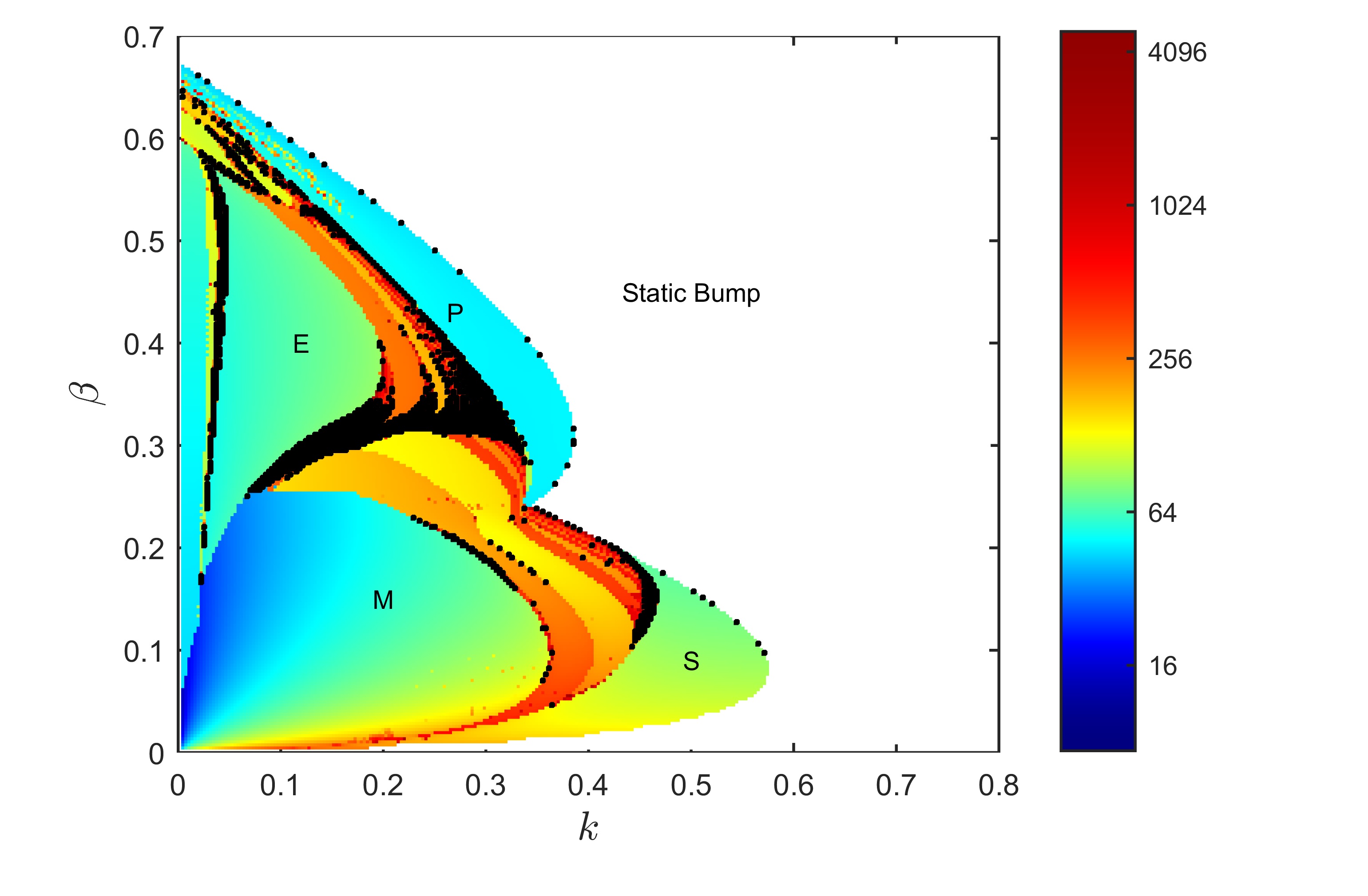}
\caption{\label{fig:pda08}
(Color online) Phase diagram in the space of $k$ and $\beta$. The color 
code indicates the lengths of the periods in log scale. Black dots means 
that the period of the response is too long to be detected by the program.
Parameters: $A=0.8$, $a=0.5$ and $a_A=\sqrt{2}/2$. 
}
\end{figure}

\begin{figure}[ht]
\includegraphics[width=8.6cm]{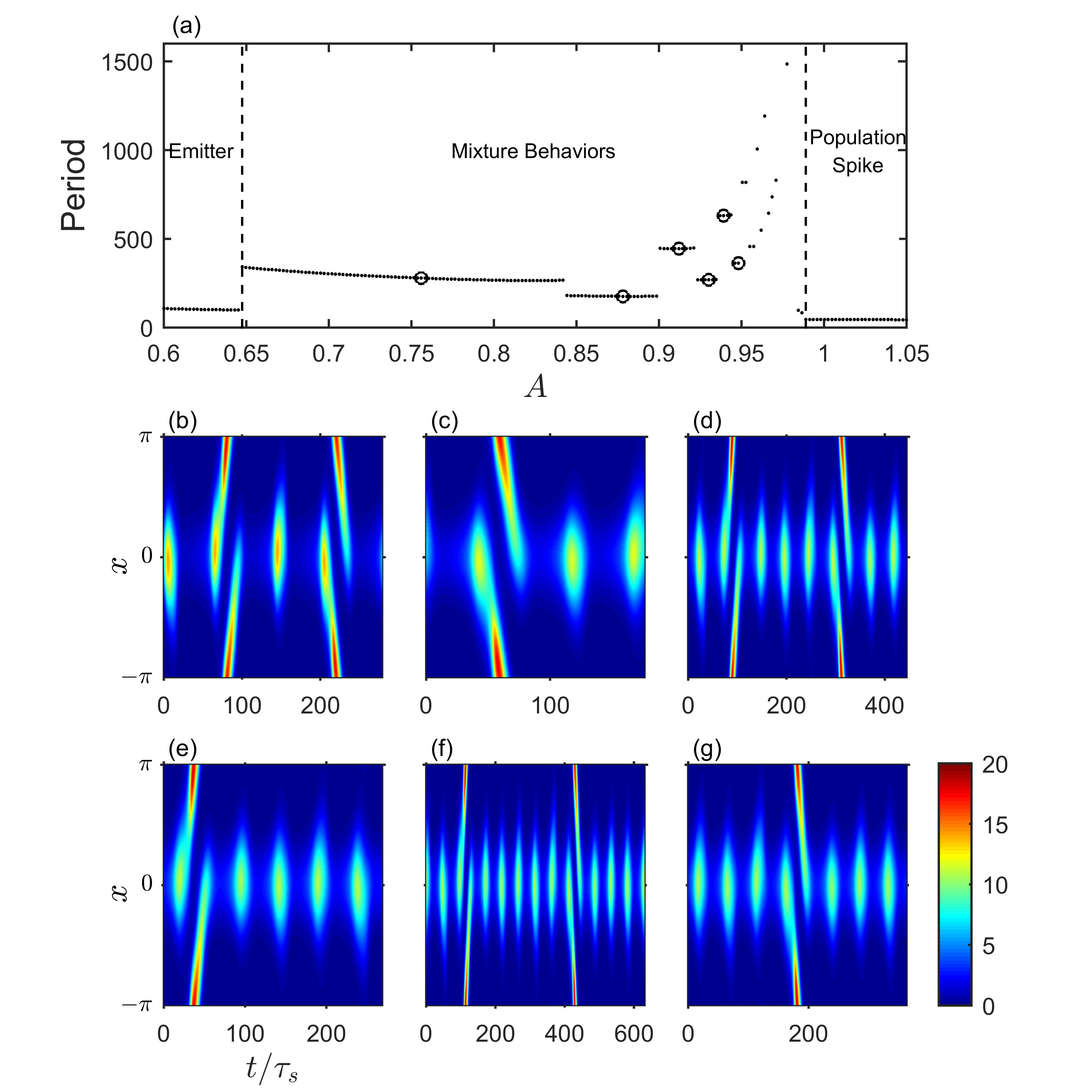}
\caption{\label{fig:scan_k03b03}
(Color online) Mixtures of emitters and population spikes. (a) Periods of 
different responses with different input strength $A$.
Dots are simulation results. If the period is too long to be 
detected, there is no dot. Six open circles are places where examples in 
(b)-(g) are drawn from. (b)-(g) Examples of mixture behaviors between 
emitters and population spikes. They correspond to the circles in (a) from 
left to right. Color scale indicates the firing rate $r(x,t)$. Each of (b)-(g) 
shows one period of the particular behavior.
$k=0.3$, $\beta=0.3$, and $a=a_A=0.8378$.
}
\end{figure}

\begin{figure}[ht]
\includegraphics[width=8.6cm]{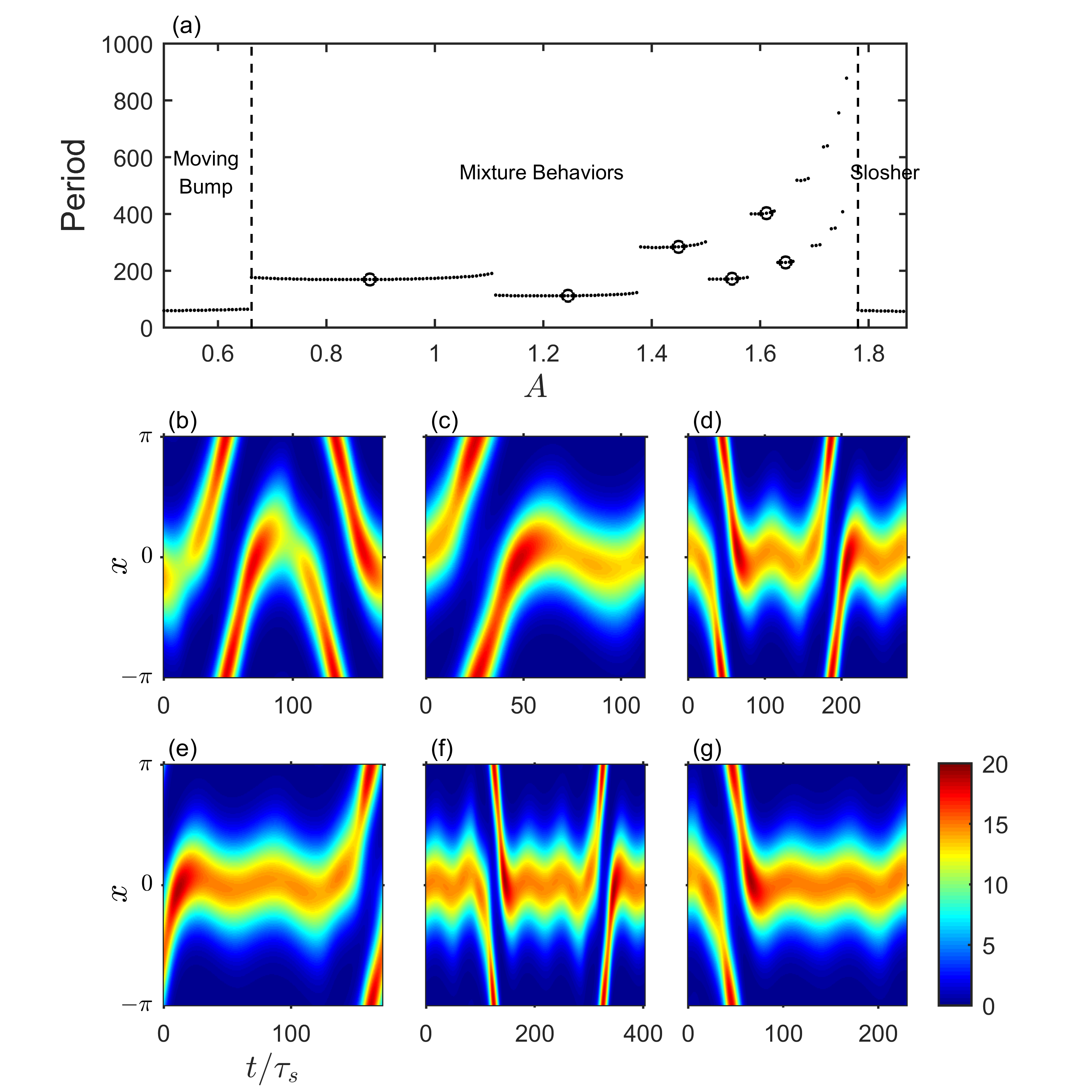}
\caption{\label{fig:scan_k03b013}
(Color online) Mixtures of moving bumps and sloshers. (a) Periods of 
different responses with different input strength $A$.
Dots are simulation results. If the period is too long to be 
detected, there is no dot. Six open circles are places where examples in 
(b)-(g) are drawn from. (b)-(g) Examples of mixture behaviors between 
moving bumps and sloshers. They correspond to the circles in (a) from left 
to right. Color scale indicates the firing rate $r(x,t)$. Each of (b)-(g) shows 
one period of the particular behavior.
$k=0.3$, $\beta=0.13$, and $a=a_A=0.8378$.
}
\end{figure}

\begin{figure}[ht]
\includegraphics[width=8.6cm]{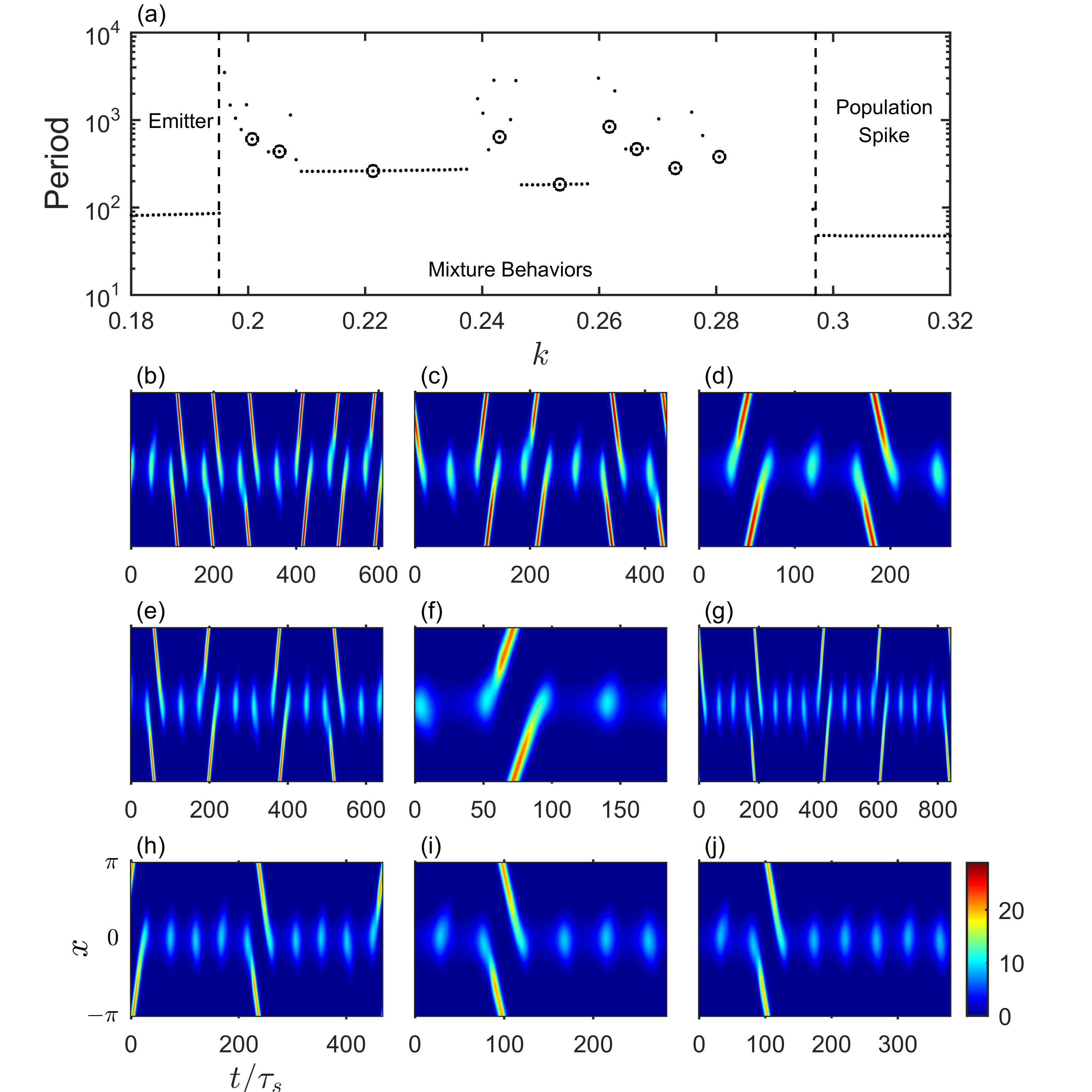}
\caption{\label{fig:scan_a08b036}
(Color online) Mixtures of emitters and population spikes. (a) Periods of 
different responses with different global inhibition strength $k$. The 
vertical axis is in logarithmic scale. Dots are simulation results.
If the period is too long to be detected, there is no 
dot. Nine open circles are places where examples in (b)-(j) are drawn from. 
(b)-(j) Examples of mixture behaviors between emitters and population spikes. 
They correspond to the circles in (a) from left to right. Color scale 
indicates the firing rate $r(x,t)$. Each of (b)-(j) shows one period of the 
particular behavior.
$\beta=0.36$, $A=0.8$, $a=0.5$ and $a_A=\sqrt{2}/2$.
}
\end{figure}

\begin{figure}[ht]
\includegraphics[width=8.6cm]{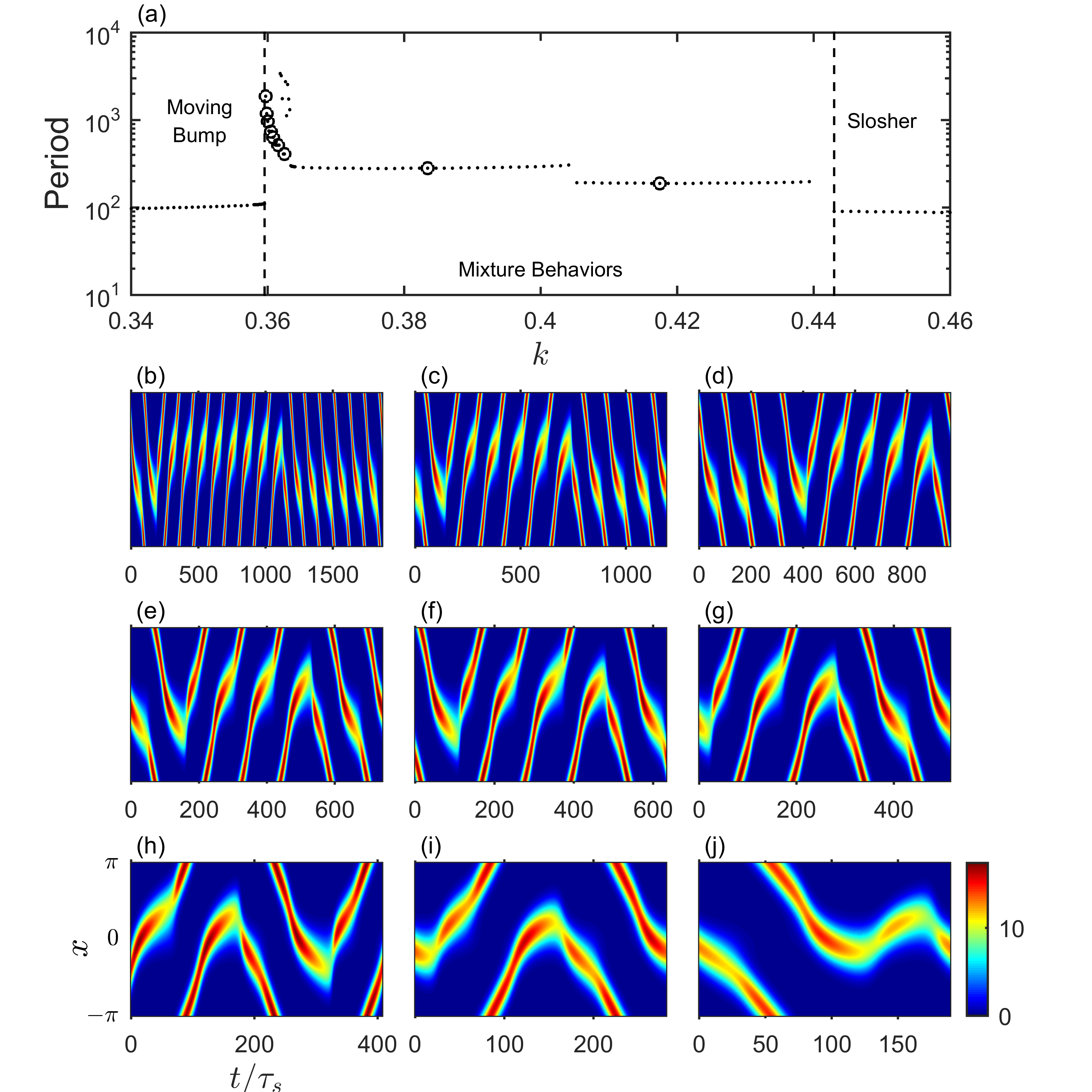}
\caption{\label{fig:scan_a08b01}
(Color online) Mixtures of moving bumps and sloshers. (a) Periods of 
different responses with different global inhibition strength $k$. The 
vertical axis is in logarithmic scale. Dots are simulation results. 
If the period is too long to be detected, there is no 
dot. Nine open circles are places where examples in (b)-(j) are drawn from. 
(b)-(j) Examples of mixture behaviors between moving bumps and sloshers. 
They correspond to the circles in (a) from left to right. Color scale 
indicates the firing rate $r(x,t)$. Each of (b)-(j) shows one period of the 
particular behavior.
$\beta=0.1$, $A=0.8$, $a=0.5$ and $a_A=\sqrt{2}/2$.
}
\end{figure}

In both cases of the moving bump and the slosher, the dynamics are governed by 
the mobility enhanced by the STD and 
the attraction provided by the external input. 
The amplitude of the bump does not change significantly during its movement. 
However, when $\beta$ and $A$ are sufficiently large, the effect of the STD is 
not just mobility enhancement, but also amplitude modulation. Large $\beta$ 
and $A$ means that the bump will consume more neurotransmitters so that it 
cannot maintain its amplitude all the time. 
The emitter (Fig.~\ref{fig:4basic}(a)) is an example in such case. 
One moving bump is emitted by the external input. 
After it travels around the network, the bump dies down due to the excessive 
consumption of neurotransmitters during traveling. Then, the network waits a 
while until sufficient amount of neurotransmitters is recovered to support 
another emission of the moving bump. When the external input is even 
stronger, we see a similar response, namely, population spikes 
(Fig.~\ref{fig:4basic}(b)), in which case a static bump, rather than a branch
of moving bump, is emitted after recovery, since the external input is 
so strong that the bump is trapped \cite{Loebel2002,Mark2012}.
This behavior is similar to breathers
\cite{Kilpatrick2010,Coombes2005,Folias2005,Folias2004}, except that 
breathers oscillate in their widths, whereas population spikes primarily
oscillate in their heights.

We explore the four basic dynamic responses in the parameter space of $k$ and 
$\beta$ with different values of $A$ (Fig.~\ref{fig:PD_As}). When the input 
is weak (Fig.~\ref{fig:PD_As}(a) and (b)), sloshers appear between the 
moving bump region and the static bump region, compared to the intrinsic 
behaviors in Fig.~\ref{fig:intrin_pd}(a). 
This is because the mobility enhanced by the STD is not enough to 
delocalize the bump, which is attracted by the static input.
When the input is strong (Fig.~\ref{fig:PD_As}(c) and (d)), 
the consumption of neurotransmitters 
is so fast that the bump cannot keep its amplitude stable. This results 
in the emergence of the emitter and population spikes. 
We also notice that when the input is weak, 
there is a bistable region for static bumps.
In this bistable region, there are two stable static bump solutions,
corresponding to the self-sustained bump and a weaker bump that is 
created only when $A>0$.
A bifurcation diagram of these static bump solutions is shown in 
Fig.~\ref{fig:si5_conti_cusp} and discussed in Appendix \ref{sct:fs}.

\subsection{Mixture Behaviors}

In numerical solutions, we find that in a large part of the parameter 
space, response patterns can be none of the four basic patterns and very 
complex. They seem to be different mixtures of the four basic dynamic patterns. 
Most of these responses are periodic. The temporal duration of one period 
is closely related to what kind of responses are being mixed. Similar 
behaviors have been observed in other models of neural fields with spike 
frequency adaptation or short-term synaptic depression 
\cite{Folias2004,Folias2005}. However, relations between different mixture 
behaviors have not been systematically understood. Here, we proposed that 
this can be done by monitoring the period of asymptotic states. Thus, we can 
show how different mixture behaviors 
are organized in the phase diagram in the space of $A$ and $\beta$ 
(Figs.~\ref{fig:pdk03_compare} and \ref{fig:pd_kkk}), and in the space 
of $k$ and $\beta$ (Fig.~\ref{fig:pda08}).

For the full model simulation (Fig.~\ref{fig:pdk03_compare}(a)), where the 
number of neurons is $256$, Eqs.~(\ref{eq:ur})-(\ref{eq:rr}) are solved by 
using the MATLAB command ode45 and the period is determined by 
examining the auto-correlation function of asymptotic states. In 
Fig.~\ref{fig:pdk03_compare}(b), the period is determined by solving a 
boundary value problem of the second order Fourier series expansion 
(see Eq.~(\ref{eq:fs}) in Appendix \ref{sct:fs}) while letting $A$ 
increases or decreases, using the MATLAB command bvp4c.

In the phase diagram, there are many patches within which the period of 
the dynamics changes continuously. However, the period jumps abruptly 
across boundaries of the patches. This indicates that behaviors are similar 
within each patch, while transitions happen across boundaries. Boundaries 
are relatively coarse in Fig.~\ref{fig:pdk03_compare}(a) due to numerical 
errors and the bi-stability of the dynamics, which is shown clearly in 
Fig.~\ref{fig:pdk03_compare}(b).
In the gray region along phase boundaries, different dynamics can be found 
by starting with initial conditions from different sides of the boundaries. 
The four basic dynamic response patterns are located at the four 
disjoint regions of the phase diagram. In between them, there is a rich 
spectrum of different mixture behaviors. Especially, black dots, where the 
length of the period is too long to be well determined within a time limit, 
are found in the mixture behavior region, which may imply chaos. Largest 
Lyapunov exponents are computed using Wolf's algorithm 
\cite{Wolf1985,Sprott2003} in regions containing black dots 
(Fig.~\ref{fig:lle}). Positive exponents exist extensively around 
$\beta=0.2$, which clearly demonstrates the existence of chaos in this 
strongly coupled neural field. 
In the green area to the left of the sloshers region,
sloshers are quasi-periodic due to the Neimark-Sacker bifurcation \cite{EABT3}
illustrated in Fig.~\ref{fig:si5_conti_cusp} in Appendix \ref{sct:fs}.
The similarity between 
Fig.~\ref{fig:pdk03_compare}(a) and (b) justifies the method of using the
Fourier basis to study this system.

\subsection{Phase Diagrams}

Having explored the phase diagrams in different situations, we find that the 
parameter space can be separated into two parts. The upper part where 
$\beta$ is larger consists of emitters, population spikes and their mixtures 
(Fig.~\ref{fig:scan_k03b03} and \ref{fig:scan_a08b036}). The lower part 
where $\beta$ is smaller consists of moving bumps, sloshers and their mixtures 
(Fig.~\ref{fig:scan_k03b013} and \ref{fig:scan_a08b01}). Between these two 
parts around $\beta=0.2$, responses tend to have very long period and show 
chaotic features. We may conclude that short-term synaptic depression have 
different effects depending on its strength. 
 
Weak STD enhances the mobility of the bump without affecting the amplitude 
of the bump significantly, leading to bumps of relatively stable amplitude 
and varying position. This is the spatial modulation effect of the STD. In 
Fig.~\ref{fig:scan_a08b01}, we see that as the external input gets stronger and 
stronger, the mixture behaviors tend to have less and less emitter components 
but more and more slosher components, from Fig.~\ref{fig:scan_a08b01}(b) to (g).

On the other hand, strong STD disrupts the bump in time, since 
neurotransmitters are depleted rapidly during the spikes. This shows the 
temporal modulation effect of the STD. Bumps in the time sequence generally are 
not the same, implying a possibility to encode different information
in different emissions, an example having been discussed in detail
in \cite{Fung2013}. In Fig.~\ref{fig:scan_k03b03}, we see that emission of 
population spikes progressively dominate over emission of moving bumps 
as $A$ increases from Fig.~\ref{fig:scan_k03b03}(b) to (g).

Varying the global inhibition at a constant external stimulus strength shows 
similar effects. Stronger inhibition suppresses activities in the region 
outside the external stimulus, therefore confines the bump or spikes to 
the region near the external stimulus. Hence for the case of strong STD 
shown in Fig.~\ref{fig:scan_a08b036}, the emitter components in the 
mixture behavior are progressively replaced by population spikes when 
global inhibition increases. Similarly, for the case of weak STD in 
Fig.~\ref{fig:scan_a08b01} the moving bump components in the mixture are 
progressively replaced by sloshers when global inhibition increases.

\section{\label{sec:discussion}discussion}
We have found a rich spectrum of firing patterns in CANNs with STD in the 
regime of weak inhibition and the regime of a single static input. CANNs with 
moderately strong inhibition were initially introduced to track continuous 
inputs using static and moving bumps. However, in the very weak global 
inhibition region, CANNs can no longer support bumps. Instead, the dynamics 
of the network show population spikes of various kinds. In particular, chaotic 
behavior is found in the amplitudes of the population spikes, typical of 
cycles of storing and releasing resources (neurotransmitters) in pulses. The 
smallness of STD scales as $\tau_s/\tau_d$ and the smallness of the global 
inhibition scales as $(\tau_s/\tau_d)^2(a/L)$.

In the case of a single static input, we have found four basic patterns of 
dynamic responses and their mixtures.
When STD is weak (lower than $0.2$, roughly), 
it mainly provides spatial modulation, or in other words, enhances
the mobility of the bump. Inputs of different strengths 
provide different attraction, leading to moving bumps, 
sloshers or mixtures of them. When STD is strong, STD provides temporal 
modulation along with spatial modulation. Together with the static external 
input, they results in emitters, population spikes, or mixtures of them. In the 
parameter region where STD strength is intermediate, chaotic behaviors appear. 
Although it is not fully understood, we believe that the involvement of 
both temporal and spatial modulation of STD is the major cause of 
complexity in that region.

Due to their richness, the firing patterns have potentials in encoding and 
decoding information. An example is the decoding of two inputs that fluctuate in
time and overlap in the space of preferred stimuli so strongly that their 
time average becomes indistinguishable. Hence any time-independent decoding 
methods are rendered ineffective. In \cite{Fung2013} we demonstrated that 
temporal modulations of the population spikes can provide a mechanism to 
resolve the two inputs, and produce results consistent with the resolution 
enhancement in transparent motion \cite{Treue2000}.

Bifurcation analysis has provided important insights into the neural field
dynamics, indicating underlying mechanisms for a variety of dynamical behaviors.
In \cite{Rankin2013}, a similar neural field model with linear spike frequency
adaptation was studied with detailed bifurcation analysis. In the presence of a 
simple weak input, their model also shows static bumps, moving bumps, and 
sloshers. The population spikes, which need a relatively strong input 
(Fig.~\ref{fig:PD_As}), were not found in \cite{Rankin2013} 
because the inputs were relatively weak. A potential mechanism for perception
switching with complex inputs mediated by sloshers was proposed.
In \cite{Cortes2013}, a mechanism leading to chaos was shown in effect in the 
single neuron model with short-term synaptic plasticity. The chaotic behavior
emerges through a Shil'nikov bifurcation of homoclinic orbits. Together with 
our mechanism on the population level, how the short-term synaptic plasticity
enhances signal processing in the neural system across
multiple levels should be of interest for further studies.

Furthermore, the existence of chaotic behaviors in our network is relevant to 
the so-called ``edge-of-chaos'' region, which has been observed to coincide with
best computational competence 
\cite{PhysicaD_1990_Langton, NeCo_2004_Bertschinger}.
Near the edge of chaos, the behavior of the network is neither dominated by 
the internal dynamics so as to be insensitive to external inputs, nor does it 
depend on the external perturbations so much as to be vulnerable to any noise. 
Although some argued that operating near the edge of chaos is 
neither a sufficient nor necessary condition for the system to achieve 
the best computational power \cite{NeurNet_2007_Legenstein}, 
chaotic neural networks are still often resorted to 
in modeling generic cortical microcircuits \cite{NRN_2009_Buonomano}.

Our work shows that even a recurrent network with a highly regular structure
can support extremely complex dynamics and chaos, in the presence of 
short-term synaptic plasticity. Previous work showed that when randomly 
connected recurrent neural networks exhibit chaotic activities, they act as 
a dynamical repertoire powerful in performing a variety of complex computational
tasks \cite{Neuron_2009_Sussillo, CC_2014_Hoerzer} and capable of reproducing 
main features of certain experiments \cite{Nature_2013_Mante} through different
learning rules. However, the randomness in their connectivity and the 
overwhelming richness of their dynamics make theoretical understanding and 
predictable generalization difficult \cite{JCSS_2004_Maass, NeurNet_2013_Ju},
although in some cases dynamical skeletons can be extracted to elucidate 
how the network achieves different functions \cite{COIN_2014_Sussillo}.
In CANNs with STD, the edge of chaos emerges in the region where the primary 
effect of STD changes from spatial modulation to temporal modulation. 
A rich spectrum of dynamical behaviors can be readily found and understood near 
the edge of chaos.  How to tap into the potential computational power of CANNs 
with STD is an interesting problem to be investigated in the future.

\begin{acknowledgments}
This work is supported by the Research Grants Council of Hong Kong 
(grant numbers 604512, 605813 and N\_HKUST606/12),
National Basic Research Program of China (2014CB846101) and 
the National Natural Science Foundation of China (31261160495).
We are grateful to Jean-Pierre Nadal and Gianluigi Mongillo 
for helpful discussions.
\end{acknowledgments}

\appendix
\section{\label{sct:apd_a}
stability analysis of fixed points in the simplified homogeneous dynamics}
The trace $T$ and determinant $D$ of the Jacobian matrix of 
Eqs.~(\ref{eq:du_uniform}) and (\ref{eq:dp_uniform}) are,
\begin{eqnarray}
T&=&-1+J_a\frac{2p_0u_0}{B^2}-\frac{\tau_s}{\tau_d}\left(1+\frac{\beta u_0{}^2}{B}\right),
\label{eq:homo_tr}\\
D&=&\frac{\tau_s (u_0J_a-2)}{\tau_d B}.
\label{eq:homo_det}
\end{eqnarray}

For the two fixed point solutions in Eq.~(\ref{eq:u_fixed_uniform}),
\begin{equation}
D=\pm\frac{\tau_s\sqrt{J_a^2-4\gamma}}
{2\gamma\tau_d B}\left(J_a\pm\sqrt{J_a^2-4\gamma}
\right),
\label{eq:homo_det_pm}
\end{equation}
where $\gamma\equiv\beta+kL/(8\sqrt{2\pi}a)$ and plus signs are for the 
solution with larger $u$. $D$  is the product of the two eigenvalues. 
Therefore, the fixed point with smaller $u$ is a saddle point with 
$D<0$ (triangles in Fig.~\ref{fig:limit_cycle}), whereas the stability of 
the fixed point with larger $u$ ($D>0$) depends on the sign of the trace 
$T$. Combining Eq.~(\ref{eq:homo_tr}) with the expression for the 
nullclines in Eq.~(\ref{eq:homo_nul}) and the definition for $B$, we can 
derive the parametric expression (Eq.~(\ref{eq:homo_hopf})) for the 
boundary where $T=0$.

\section{\label{sct:fs}Fourier series expansion of the CANNs model}
For the CANN model Eqs.~(\ref{eq:ur})-(\ref{eq:rr}) and the external input 
$I(x,t)=A\exp [-{(x-z)^{2}}/({2a_A^{2}})]$, where $a_A$ is the width of the
 input, we expand $U(x,t)$ and $p(x,t)$ in terms of Fourier series up to 
$M^{th}$ order,
\begin{eqnarray}
U(x,t)&=&\sum_{l=-M}^{M}{u_l(t)e^{i2\pi lx/L}},\nonumber\\
p(x,t)&=&\sum_{l=-M}^{M}{p_l(t)e^{i2\pi lx/L}},\label{eq:ft_expansion}
\end{eqnarray}
where,
\begin{eqnarray}
u_l(t)&=&\frac{1}{L}\int_{-L/2}^{L/2}{u(x,t)e^{-i2\pi lx/L}dx},\nonumber\\
p_l(t)&=&\frac{1}{L}\int_{-L/2}^{L/2}{p(x,t)e^{-i2\pi lx/L}dx}.\label{eq:ft_coef}
\end{eqnarray}

\begin{figure}[b]
\includegraphics[width=8.6cm]{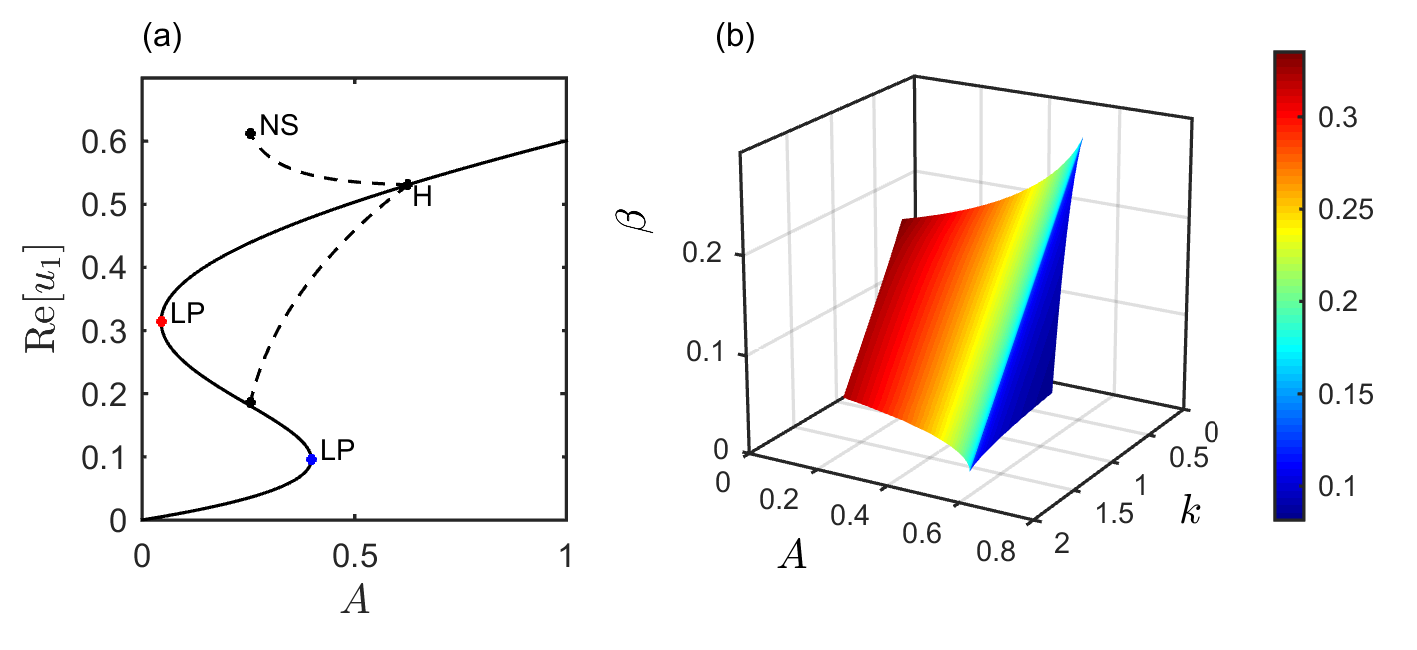}
\caption{\label{fig:si5_conti_cusp}
(Color online) (a) The real part of the first order component $u_1$
of solutions of Eq.~(\ref{eq:fs}) with different strengths of 
external inputs $A$. The solid line is fixed point solutions.
The red dot and blue dot labeled LP (limit points) corresponds to 
two saddle-node bifurcations, respectively. 
The black dot labeled H indicates a Hopf bifurcation.
The dashed lines are extrema of limit cycles.
The black dot labeled NS indicates a Neimark-Sacker bifurcation.
Other parameters: $k=0.55$, $\beta=0.08$, $a=0.5$ and $a_A=\sqrt{2}/2$.
(b) Continuation of the saddle-node bifurcation in the parameter space. 
The left and right surfaces corresponds to the red and blue dots 
in panel (a), respectively. The color scale shows 
the real part of the first order component $u_1$.
The cusp bifurcation happens along the intersection of the two surfaces.
\label{fig:cusp}
}
\end{figure}

Therefore, the CANN model can be rewritten as,
\begin{eqnarray}
\tau_s\frac{\partial u_l(t)}{\partial t}&=&-u_l(t)+A\frac{\sqrt{2\pi}a_A}{L}e^{-2a_A^2\pi^2 l^2/L^2}\nonumber\\
&&+\frac{1}{B(t)}e^{-2a^2\pi^2 l^2/L^2}\sum_{q-s+v=l}{p_q(t)u_s^*(t)u_v(t)},\nonumber\\
\tau_d\frac{\partial p_l(t)}{\partial t}&=&\delta_{l,0}-p_l(t)-\frac{\beta}{B(t)}\sum_{q-s+v=l}{p_q(t)u_s^*(t)u_v(t)},\nonumber\\
B(t)&=&1+\frac{kL}{8\sqrt{2\pi}a}\sum_{l=-M}^{M}{u_l^*(t)u_l(t)}.\label{eq:fs}
\end{eqnarray}
where $\delta_{l,0}$ is equal to $1$ only when $l=0$, otherwise $0$. 
$u_s^*(t)$ is the complex conjugate of $u_s(t)$. $q$, $s$, $v$ are 
all integer indices ranging from $-M$ to $M$.

For $M=0$, Eq.~(\ref{eq:fs}) reduces to the simplified homogeneous dynamics 
in Eqs.~(\ref{eq:du_uniform}) and (\ref{eq:dp_uniform}). For $M=1$, 
Eq.~(\ref{eq:fs}) is equivalent to Eq.~(\ref{eq:wave_linearization}) where the 
wave stability is analyzed. For $M=2$, it would be  complicated to apply  
analytical methods. However, numerical methods are far more efficient for 
the Fourier series expansion up to the second order than for the full network 
model. Moreover, almost all features of the phase diagram of the full network 
model are maintained in the phase diagram of Eq.~(\ref{eq:fs}) with $M=2$ 
(see Fig.~\ref{fig:pdk03_compare}(a) and (b)).

A bifurcation diagram is computed numerically using MATCONT \cite{Matcont2003} 
to illustrate the bistable region in Fig.~\ref{fig:PD_As} and the emergence of 
sloshers. Within the region in Fig.~\ref{fig:si5_conti_cusp}(b), 
there are three static bump solutions centered at $x=0$.
The lower one and higher one corresponds to 
the weak bump that is created only when $A>0$ and the self-sustained bump.
The middle one is a saddle point
which separates basins of attraction of the other two bump solutions.
The slosher appears after a Hopf bifurcation of the higher static bump state.
When $A$ decreases, it will become unstable after a Neimark-Sacker bifurcation 
\cite{EABT3}, which produces quasi-periodic behaviors shown in the green area
to the left of  the sloshers region in Fig.~\ref{fig:lle}.

\bibliography{hewang_2014}
\end{document}